\documentclass[twocolumn,superscriptaddress,footinbib,amsmath,amssymb,aps,prx,letterpaper,10pt,longbibliography,floatfix]{revtex4-2}

\usepackage{graphicx}% Include figure files
\usepackage{array,siunitx}
\usepackage{soul}
\usepackage{bm,physics}% bold math
\usepackage[utf8]{inputenc}
\usepackage[english]{babel}
\usepackage[dvipsnames]{xcolor}
\usepackage{dcolumn}
\usepackage[colorlinks=true, urlcolor=blue, pdfborder={0 0 0}]{hyperref}% add hypertext capabilities

\newcommand{\req}[1]{Eq.~(\ref{#1})}
\newcommand{\reqs}[1]{Eqs.~(\ref{#1})}
\newcommand{\rref}[1]{(\ref{#1})}

\newcommand{\sto}{\mathrm{SrTiO}{}_{3}}

\newcommand{\unitAlpha}{\mathrm{J} \, \mathrm{m} \, \mathrm{C}^{-2}}
\newcommand{\unitBeta}{\mathrm{J} \, \mathrm{m}^{5} \, \mathrm{C}^{-4}}

\begin{document}
%\preprint{APS/123-QED}
\title{Kapitza stabilization of quantum critical order}%.~\noteDK{REVISION 2}
%AUTHOR 1
\author{Dushko Kuzmanovski}
\affiliation{
Nordita, KTH Royal Institute of Technology and Stockholm University
Hannes Alfv\'{e}ns v\"{a}g 12, SE-106 91 Stockholm, Sweden
}
% AUTHOR 2
\author{Jonathan Schmidt}
\affiliation{Department of Materials, ETH Z\"{u}rich, Z\"{u}rich, CH-8093, Switzerland}
% AUTHOR 3
\author{Nicola A. Spaldin}
\affiliation{Department of Materials, ETH Z\"{u}rich, Z\"{u}rich, CH-8093, Switzerland}
% AUTHOR 4
\author{Henrik M. R\o{}nnow}
\affiliation{Laboratory for Quantum Magnetism, Institute of Physics,
\'{E}cole Polytechnique F\'{e}d\'{e}rale de Lausanne, CH-1015 Lausanne, Switzerland}
% AUTHOR 5
\author{Gabriel Aeppli}
\affiliation{Paul Scherrer Institut, Villigen, PSI CH-5232, Switzerland}
\affiliation{Institut de Physique, EPFL, Lausanne, CH-1015, Switzerland}
\affiliation{Department of Physics and Quantum Center, ETH Z\"{u}rich, Z\"{u}rich, CH-8093, Switzerland}
% AUTHOR 6
\author{Alexander V. Balatsky}
\affiliation{
Nordita, KTH Royal Institute of Technology and Stockholm University
Hannes Alfv\'{e}ns v\"{a}g 12, SE-106 91 Stockholm, Sweden
}
\affiliation{
Department of Physics, University of Connecticut, Storrs, Connecticut 06269, USA
}
\date{\today}% It is always \today, today,

\begin{abstract}
Dynamical perturbations modify the states of classical systems in surprising ways and give rise to important applications in science and technology. For example, Floquet engineering exploits the possibility of band formation in the frequency domain when a strong, periodic variation is imposed on parameters such as spring constants. We describe here Kapitza engineering, where a drive field oscillating at a frequency much higher than the characteristic frequencies for the linear response of a system changes the potential energy surface so much that maxima found at equilibrium become local minima, in precise analogy to the celebrated Kapitza pendulum where the unstable inverted configuration, with the mass above rather than below the fulcrum, actually becomes stable. Our starting point is a quantum field theory of the Ginzburg-Devonshire type, suitable for many condensed matter systems, including particularly ferroelectrics and quantum paralectrics.
%such as the common substrate (for oxide electronics) strontium titanate ($\sto$).
We show that an off-resonance oscillatory electric field generated by a laser-driven THz source can induce ferroelectric order in the quantum-critical limit. Heating effects are estimated to be manageable using pulsed radiation; ``hidden" radiation-induced order can persist to low temperatures without further pumping due to stabilization by strain. We estimate the Ginzburg-Devonshire free-energy coefficients in $\sto$ using density-functional theory (DFT) and the stochastic self-consistent harmonic approximation accelerated by a machine-learned force field. Although we find that $\sto$ is not an optimal choice for Kapitza stabilization, we show that scanning for further candidate materials can be performed at the computationally convenient density-functional theory level. We suggest second-harmonic-generation, soft-mode-spectroscopy, and X-ray-diffraction experiments to characterize the induced order.
\end{abstract}

\maketitle

\section{\label{sec:Intro}Introduction}
The Kapitza effect~\cite{KapitzaZhETF1951, *1965714, KapitzaUspFN1951, *1965726, Landau1982mechanics} takes place when an anharmonic oscillator is dynamically stabilized at unstable equilibria by driving it at a frequency significantly higher than its natural frequency. The motion is divided into slow and fast components, with time averaging over the fast drive's short period enabling the analysis of the slow-mode motion's stability in terms of an effective potential that depends on the drive's amplitude and frequency. The concept that fast degrees of freedom can modify slow behavior is not new, as illustrated by the thermal expansion of solids induced by rapid lattice fluctuations. This extreme rectification of fast fluctuations into a slowly varying (dc) behavior embodies the Kapitza engineering idea.

Kapitza engineering  has been employed to stabilize repulsive Bose-Einstein condensates in an optical lattice~\cite{MartinPRA2018, *MartinPRA2018Err}, periodically driven spin systems~\cite{LerosePRB2019}, a many-body generalization of the Kapitza pendulum as a periodically driven sine-Gordon model~\cite{CitroAnnPhys2015}, and a Josephson junction coupled to a nanomagnet~\cite{KulikovPRB2022}. In each instance, the system is characterized by a nonlinear equation of motion for an effective degree of freedom coupled to a generalized field that induces externally controlled fast oscillations.

In this work, we suggest using Kapitza stabilization to dynamically control Quantum Matter near a quantum critical point (QCP). To develop this proposal, we use a model for the QCP state. As a prototypical example, we consider incipient ferroelectric (FE) materials like $\sto$ (STO)\cite{MuellerPRB1979, RowleyNatPhys2014, EdgePRL2015}, where the QCP can be tuned by control parameters such as strain\cite{HaeniNat2004, VermaAPL2015}, Ca~\cite{BednorzPRL1984, RischauNatPhys2017}, and ${}^{18}$O isotope doping~\cite{ItohPRL1999}. Incipient displacive FEs offer several advantages, including the availability of laser-driven electric fields $\vb{E}(\vb{r},t)$ directly coupled to the polarization. Additionally, the typical time scales associated with optical phonons responsible for displacive polarization lie within the femtosecond to picosecond range. We can therefore implement the Kapitza approach by applying fields in the accessible few-terahertz regime, as discussed below.

Dynamic control of FEs using infrared laser-driven radiation represents a broad approach to terahertz manipulation of quantum matter~\cite{RiniNat2007, FoerstNatPhys2011, QiPRL2009, NovaNatPhys2017, SubediPRB2017, KozinaNat2019, NovaSci2019, LiSci2019, SalenPhysRep2019, ShinPRL2022, DisaNaPhys2021}, including Floquet engineering~\cite{BukovAiP2015, OkaARCMP2019, TindallPRB2021, TindallQuantum2021}. The goal is to resonantly couple the photon to an optically-active phonon mode~\cite{KozinaNat2019, NovaSci2019, LiSci2019} for the most optimal energy transfer and then exploit the anharmonic inter-ionic potential to excite lower-frequency phonon modes or cause structural lattice changes. Using resonant cavities to exploit vacuum fluctuations of hybridized phonon-photon-plasmon cavity modes has also been explored~\cite{AshidaPRX2020}. A similar effect was used in another setting to provide the pairing glue for superconductivity~\cite{SchlawinPRL2019, BuzziPRX2020, TindallPRL2020}. Besides ferroelectricity, ultrashort laser pulses have been utilized to control magnetization~\cite{Kanda2011}, and the Leggett mode in multiband superconductors~\cite{Giorgianni2019}.

Observing the relatively long-lived polarization in a nominally paraelectric state due to applied time-varying electric fields would be a significant step towards dynamic control of the quantum state. Pristine STO is a quantum paraelectric. As an ionic insulator with a strong coupling between strain and polarization, the ferroelectric state in STO can have a long lifetime after the THz radiation is turned off. The relaxation of the induced FE state would occur on longer time scales associated with strain and lattice coupling. We envisage that this coupling helps in stabilizing the induced state without energy input producing Joule heating.

Our approach to QCP under drive follows the general approach to the Kapitza pendulum. We integrate the fast harmonics of $\vb{P}$ and find that new minima of the free energy for the FE order parameter can emerge, eventually leading to instability. To make a contact with realistic parameters we use ab initio analysis for free energy parameters for paraelectric state.  In this work, we identify three essential features of the model required to induce the effect: i) a nonlinear potential, specifically the Devonshire-Landau free energy for the FE order parameter~\cite{MuellerPRB1979, AharonyPRL1977}, ii) retardation in the response, and iii) application of the drive ``off-resonance", where, as for the pendulum, the drive frequency of the applied field is chosen to be above resonance with the natural frequencies of the FE order parameter $\vb{P}$, but not so high as to induce excitation above the insulating band gap. Due to these reasons, we refer to this approach as ``Kapitza engineering", as discussed in Ref.~\onlinecite{BukovAiP2015}. 

We predict the required orientation of the optical electric field to observe the induction of QCP, contingent on the light's polarization type and the direction of the induced FE moment. Potential signatures of the induced ferroelectric order include low-frequency electrical measurements of displacement current, optical measurements of second harmonic generation, and X-ray diffraction measurements of the resulting strain.

The rest of the paper is structured as follows. In Sec.~\ref{sec:Model} we give a rather general exposition of the effective free energy of the FE and (linear) susceptibility renormalization in the presence of Kapitza drive from a strong, coherent, off-resonant light field. Then, focusing on particular crystal orientation, in Sec.~\ref{sec:Results} we analyze the critical electric field to make the PE state unstable (Sec.~\ref{subsec:PEInst}), the free-energy-gradient landscape under Kapitza drive (Sec.~\ref{subsec:FEStable}), and control of the QCP with light (Sec.~\ref{subsec:QCPcontrol}). Our ab-initio calculations to extract the GD coefficients for STO are presented in Sec.~\ref{subsec:DFTcalc}. We discuss possible detection schemes in Sec.~\ref{subsec:SecondHarmonic}, with the details of the discussion moved to the corresponding Appendix.

\section{\label{sec:Model}Model and General framework}
\begin{figure}[t]
    \centering
    \includegraphics[width=\columnwidth]{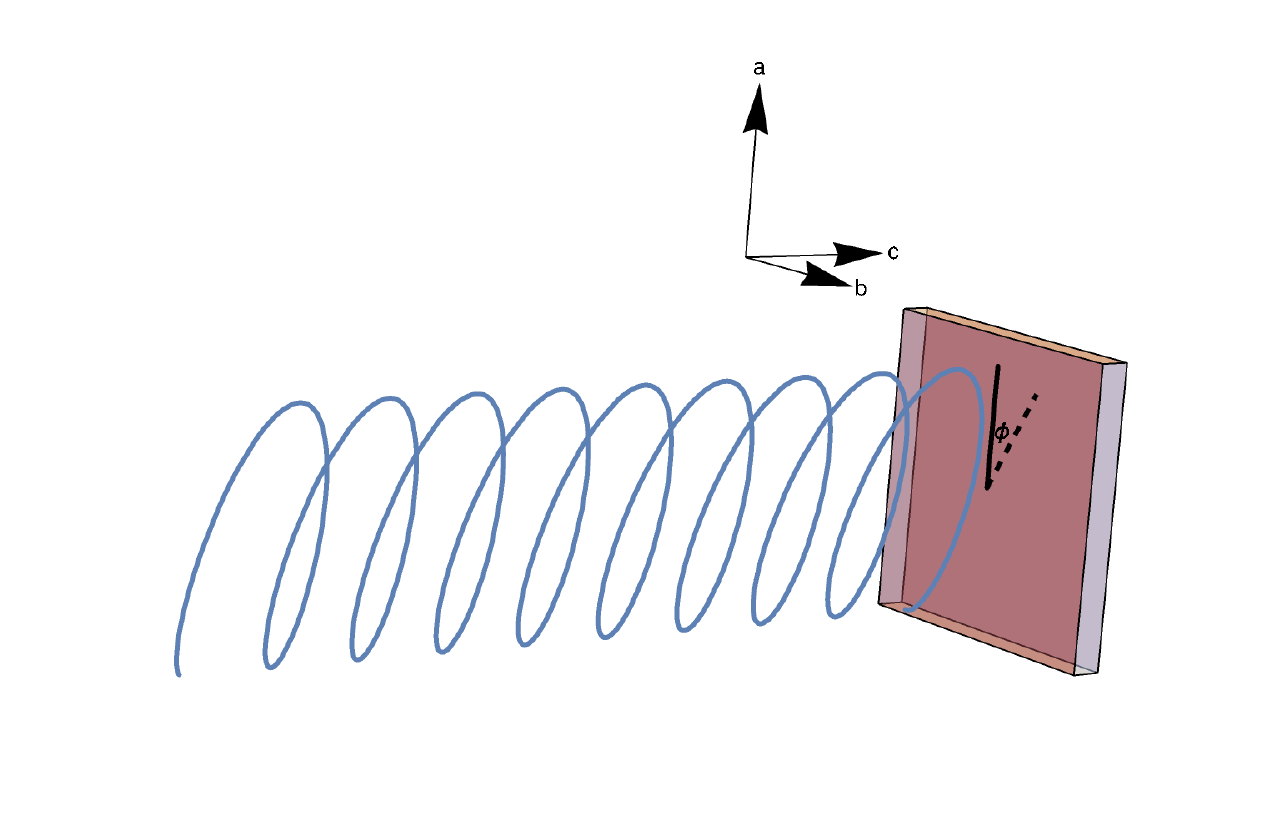}
    \caption{Diagram of the setup. A coherent laser field (blue line corresponding to  electric field vector) is incident in the $\qty[001]$ crystal direction. The electric field is in the $a-b$ crystal plane at an angle $\phi$. The incident light can have a possible polarization phase $\delta$, depicted by a spiral.}
    \label{fig:Diagram}
\end{figure}
We suppose that both the thermodynamics and the slow dynamics of the FE polarization $\vb{P}$ are governed by an effective action $\Gamma\qty[\vb{P}]$ expressed as in integral over imaginary-time $\tau$ (sum over conjugate Matsubara frequency $\nu_{n} = 2 n \, \pi/T$, where we choose units where $k_{B} = \hbar = 1$) of the free energy~\cite{Altland2010}. The effective action is composed of three terms -- a harmonic term $\Gamma_{0}$, an anharmonic term $\Gamma_{\mathrm{ah}}$, and linear coupling to the applied electric field $\vb{E}$~\cite{Landau1984}
\begin{subequations}
    \label{eq:EffAction}
    \begin{eqnarray}
    & \Gamma\qty[\vb{P}] = \Gamma_{0} + \Gamma_{\mathrm{ah}} \nonumber \\
    & - \int_{0}^{1/T} \dd{\tau} \int \dd^{3}\vb{x} \, \qty(\vb{P}\qty(\vb{x}, \tau) \cdot \vb{E}\qty(\vb{x},
    \tau)). \label{eq:EffAction1}
\end{eqnarray}
The harmonic part $\Gamma_{0}$ is given via the inverse susceptibility tensor of the unpolarized FE. Assuming translation invariance, it is best expressed in momentum-frequency space
\begin{equation}
    \frac{\varepsilon_{0} \, \Gamma_{0}}{P^{2}_{0}}= \frac{1}{2} \, \sum_{q \equiv \qty(\vb{q}, i \nu_{n})} \qty[\hat{\chi}^{M}_{q}]^{-1}_{i, j} \, \frac{\qty(\bar{P}_{-q})_{i}}{\zeta_{i}} \, \frac{\qty(\bar{P}_{q})_{j}}{\zeta_{j}}, \label{eq:HarmAction}
\end{equation}
where $\qty(\bar{P}_{q})_{j} = \sqrt{\frac{T}{L^{3}}} \, \int_{0}^{1/T} \dd{\tau} \int \dd{{}^{3}x} e^{i \, \qty(\nu_{n} \, \tau - \vb{q} \cdot \vb{x})} \, \bar{P}_{j}\qty(\vb{x}, \tau)$ is the Fourier transform with a wave-vector $\vb{q}$ and a bosonic Matsubara frequency $\nu_{n}$, and $\bar{P}_{j} = P_{j}/P_{0}$ is the dimensionless polarization scaled by some convenient unit $P_{0}$ pertaining to the crystal. Starting from cubic symmetry, we scale down to tetragonal (orthorhombic) symmetry by scaling each component $\bar{P}_{i}$ by a suitable scaling factor $\zeta_{i}$. A prototypical Ansatz for the ionic inverse susceptibility tensor is that of a Lorentzian diagonal in the orthorhombic crystal axes and with transverse propagation
\begin{eqnarray}
    & \qty[\hat{\chi}^{M}_{q}]^{-1}_{i, j} = \qty[\alpha\qty(T, g) \, - \frac{\qty(i \, \nu_{n})^{2} + 2 i \, r \, \qty(i \, \nu_{n})}{\omega^{2}_{0}}] \, \delta_{i, j} \nonumber \\
    & + \frac{c^{2}_{s}}{\omega^{2}_{0}} \, \qty[\qty(\vb{q} \cdot \vb{q}) \, \delta_{i, j} - q_{i} \, q_{j}], \label{eq:InvSuscpet1}
\end{eqnarray}
where $\alpha\qty(T, g)$ is the homogeneous, zero-frequency part, and coincides with the quadratic term of the Ginzburg-Devonshire (GD) free energy~\cite{DevonshireAdvPhys1954, KvyatkovskiiPSS2001,PertsevPRB2000,*PertsevPRB2002,RowleyNatPhys2014}. It is a function of temperature $T$, and, possibly, any parameters $g$ that control the proximity to a QCP. Since it determines the inverse square of the correlation length, the zero-temperature characteristic exponent is $2\nu$. Mean-field theory predicts $\nu = 1/2$, and other choices corresponding to a proximity to a QCP are considered in Sec.~\ref{subsec:QCPcontrol}. We model the finite temperature dependence by another non-universal exponent $n$
\begin{equation}
    \alpha\qty(T, g) \simeq \qty(\frac{c_{s} \, \Lambda}{\omega_{0}})^{2} \, \abs{\frac{g}{g_{c}} - 1}^{2 \nu} \, \mathrm{sgn}\qty(g - g_{c}) + \qty(\frac{T}{T_{0}})^{n}. \label{eq:CritExpScalingAlpha}
\end{equation}
For a classical paraelectric $n = 1$ in accordance with the Curie-Weiss law. At low temperatures, a more convenient choice is $n = 2$ corresponding to a quantum paraelectric. There are more complicated cases~\cite{CoakPNAS2020}, but this ansatz is sufficient for illustration purposes. The parameters $c_{s}$ and $r$ are the sound propagation speed and damping, respectively, and $\omega_{0}$ is a reference frequency (for example, the room-temperature value of the TO phonon frequency at the $\Gamma$ point $\omega_{\vb{q} = 0}$). The results can be extended to include any dynamical exponent $z$~\cite{sachdev_2011,ChandraRPP2017}. We choose the simplest case of a Gaussian theory with $z = 1$ for the rest of the paper.

The anharmonic action $\Gamma_{\mathrm{ah}}$ is just a space-time  integral over the (local) GD quartic free-energy density terms.
\begin{equation}
    \Gamma_{\mathrm{ah}} = \int_{0}^{1/T} \dd{\tau} \int \dd{{}^{3}\vb{x}} \mathcal{F}_{\mathrm{ah}}\qty(\vb{P}\qty(\vb{x}, \tau)), \label{eq:AnharmAction}
\end{equation}

\begin{equation}
    \frac{\varepsilon_{0} \, \mathcal{F}_{\mathrm{ah}}\qty(\vb{\bar{P}})}{P^{2}_{0}} = \frac{\beta_{1}}{4} \, \sum_{i = 1}^{3} \qty(\frac{\bar{P}_{i}}{\zeta_{i}})^{4} + \frac{\beta_{2}}{2} \, \sum_{1 \le i < j \le 3} \qty(\frac{\bar{P}_{i}}{\zeta_{i}})^{2} \, \qty(\frac{\bar{P}_{j}}{\zeta_{j}})^{2}. \label{eq:QuarticFEDens}
\end{equation}
\end{subequations}

If $\alpha < 0$, the FE would undergo a PE-FE phase transition along one of the main crystal axis. According to \req{eq:CritExpScalingAlpha}, for $g < g_{c}$, the Curie-Weiss temperature is given by $T_{CW}\qty(g) = T_{0} \, \qty(\frac{c_{s} \, \Lambda}{\omega_{0}}) \qty(1 - g/g_{c})^{2\nu/n}$. We assume the dynamics are governed by a particular optical phonon mode with a dispersion relation whose $\vb{q} = \vb{0}$ frequency is $\sqrt{\alpha} \, \omega_{0}$. Because it is a polynomial, the susceptibility \req{eq:InvSuscpet1} has a trivial analytic continuation from the experimentally measurable retarded susceptibility $\hat{\chi}^{\mathrm{ret}}_{q = \qty(\vb{q}, \omega)}$. Other Ans\"{a}tze, more suitable for proximity to a quantum critical point with an anomalous dynamic exponent $z$ may be used, but a prescription for the homogeneous $\vb{q} = 0$, zero-Matsubara-frequency term ought to be given.  The anharmonic term \req{eq:QuarticFEDens} is the most general local quartic term consistent with cubic symmetry, and the scale factors $\zeta_{i}$ account for lowered symmetry. Here, we focus on the dynamics in a plane perpendicular to the $c$-axis below the antiferrodistrotive transition~\cite{ShiranePhysRev1969,CowleySSC1969,RisteSSC1971,AschauerJPCM2014,PorerPRL2018}.

The equation of motion for the $q$-th Fourier mode is obtained by extremizing \req{eq:EffAction1}
\begin{eqnarray}
    & \frac{\varepsilon_{0} \qty(E_{q})_{i}}{P_{0}} = \qty[\chi^{M}_{q}]^{-1}_{i, j} \, \qty(\bar{P}_{q})_{j} \nonumber \\
    & + \sqrt{\frac{T}{L^{3}}} \, \int \dd{{}^{4}x} e^{-i \, \qty(q \cdot x)} \frac{\partial}{\partial \bar{P}_{i}}\qty(\frac{\varepsilon_{0} \, \mathcal{F}_{\mathrm{ah}}\qty(\vb{\bar{P}}\qty(x))}{P^{2}_{0}}). \label{eq:EqMotImTime}
\end{eqnarray}

Expanding the anharmonic term in \req{eq:EqMotImTime} up to linear order in $\bar{\vb{P}}_{q}$ around a reference static and uniform polarization $\expval{\vb{P}}$, continuing analytically from Matsubara frequencies to the upper complex half-plane in $\omega$, we obtain the effective linear susceptibility
\begin{eqnarray}
    & \qty[\chi_{\mathrm{eff}}\qty(\vb{q}, z; \expval{\vb{P}})]^{-1}_{i, j} \nonumber \\
    & = \qty[\chi_{\mathrm{eff}}\qty(\vb{q}, z)]^{-1}_{i, j} + \frac{\partial^{2}}{\partial \bar{P}_{i} \, \partial \bar{P}_{j}} \qty(\frac{\varepsilon_{0} \, \mathcal{F}_{\mathrm{ah}}\qty(\expval{\bar{\vb{P}}})}{P^{2}_{0}}). \label{eq:RenormSusc}
 \end{eqnarray}

The retarded effective susceptibility determines the real-time response under an applied electric field with components along the principal crystal axes
\begin{eqnarray}
    &\vb{E}\qty(t) = E_{0} \, \left\langle \cos\qty(\phi) \, \cos\qty(\omega \, t), \sin\qty(\phi) \, \cos\qty(\omega \, t - \delta), 0. \right\rangle \label{eq:ElField}
\end{eqnarray}
This gives rise to an oscillating component of the dimensionless polarization $\tilde{\bar{p}}_{i}$ ($\vb{\pi} = \left\langle \cos\qty(\phi),  \sin\qty(\phi) \, e^{i \, \delta}, 0\right\rangle$)
\begin{eqnarray}
    & \tilde{\bar{p}}_{i}\qty(t) = \frac{1}{2} \, \qty(\frac{\varepsilon_{0} \, E_{0}}{P_{0}}) \, \left[ e^{-i \, \omega \, t} \,  \, \chi^{\mathrm{ret}}_{\mathrm{eff}}\qty(\omega, \expval{\vb{P}})]_{i, j} \, \pi_{j} \right. \nonumber \\
    & \left. + e^{i \, \omega \, t} \, \qty[\chi^{\mathrm{ret}}_{\mathrm{eff}}\qty(-\omega, \expval{\vb{P}})]_{i, j} \, \pi^{\ast}_{j} \right]
\end{eqnarray}
The second moments over the rapidly oscillating polarization fields (remembering that $\hat{\chi}^{\mathrm{ret}}\qty(-\omega) = \qty(\hat{\chi}^{\mathrm{ret}}\qty(\omega))^{\ast}$) evaluate to
\begin{widetext}
\begin{equation}
    \expval{\tilde{\bar{p}}_{j} \, \tilde{\bar{p}}_{k}} = \frac{1}{2} \, \qty(\frac{\varepsilon_{0} \, E_{0}}{P_{0}})^{2} \, \Re\qty[\hat{\chi}^{\mathrm{ret}}_{\mathrm{eff}}\qty(\omega; \expval{\vb{P}}) \cdot \Bqty{\vb{\pi}, \vb{\pi}^{\ast}} \cdot \qty(\hat{\chi}^{\mathrm{ret}}_{\mathrm{eff}}\qty(\omega; \expval{\vb{P}}))^{\dagger}]_{j, k}. \label{eq:Secondmoment}
\end{equation}
Using \req{eq:Secondmoment}, we can express the equation of motion for the uniform and stationary solution $\expval{\vb{P}} = \sqrt{T/L^{3}} \, \vb{P}_{q = 0}$ from \req{eq:EqMotImTime} for $q = 0$ by averaging over the period of fast motion
\begin{equation}
      0 = \frac{\partial}{\partial \bar{P}_{i}} \qty(\frac{\varepsilon_{0} \, \mathcal{F}_{\mathrm{eff}}\qty(\expval{\vb{P}})}{P^{2}_{0}}) = \frac{1}{\zeta_{i}} \, \qty(\chi^{M}_{0})^{-1}_{i, j} \, \frac{\expval{\bar{P}}_{j}}{\zeta_{j}}\ + \frac{\partial}{\partial \bar{P}_{i}}\qty(\frac{\varepsilon_{0} \, \mathcal{F}_{\mathrm{ah}}\qty(\expval{\vb{P}})}{P^{2}_{0}}) + \frac{1}{2} \, \expval{\tilde{\bar{p}}_{j} \, \tilde{\bar{p}}_k} \, \frac{\partial^{3}}{\partial \bar{P}_{i} \, \partial \bar{P}_{j} \, \partial \bar{P}_{k}}\qty(\frac{\varepsilon_{0} \, \mathcal{F}_{\mathrm{ah}}\qty(\expval{\vb{P}})}{P^{2}_{0}}). \label{eq:EffFreeEnGradient}
\end{equation}
\end{widetext}

\section{\label{sec:Results}Results}
\begin{figure}[h]
    \centering
    \includegraphics[width=0.9\columnwidth]{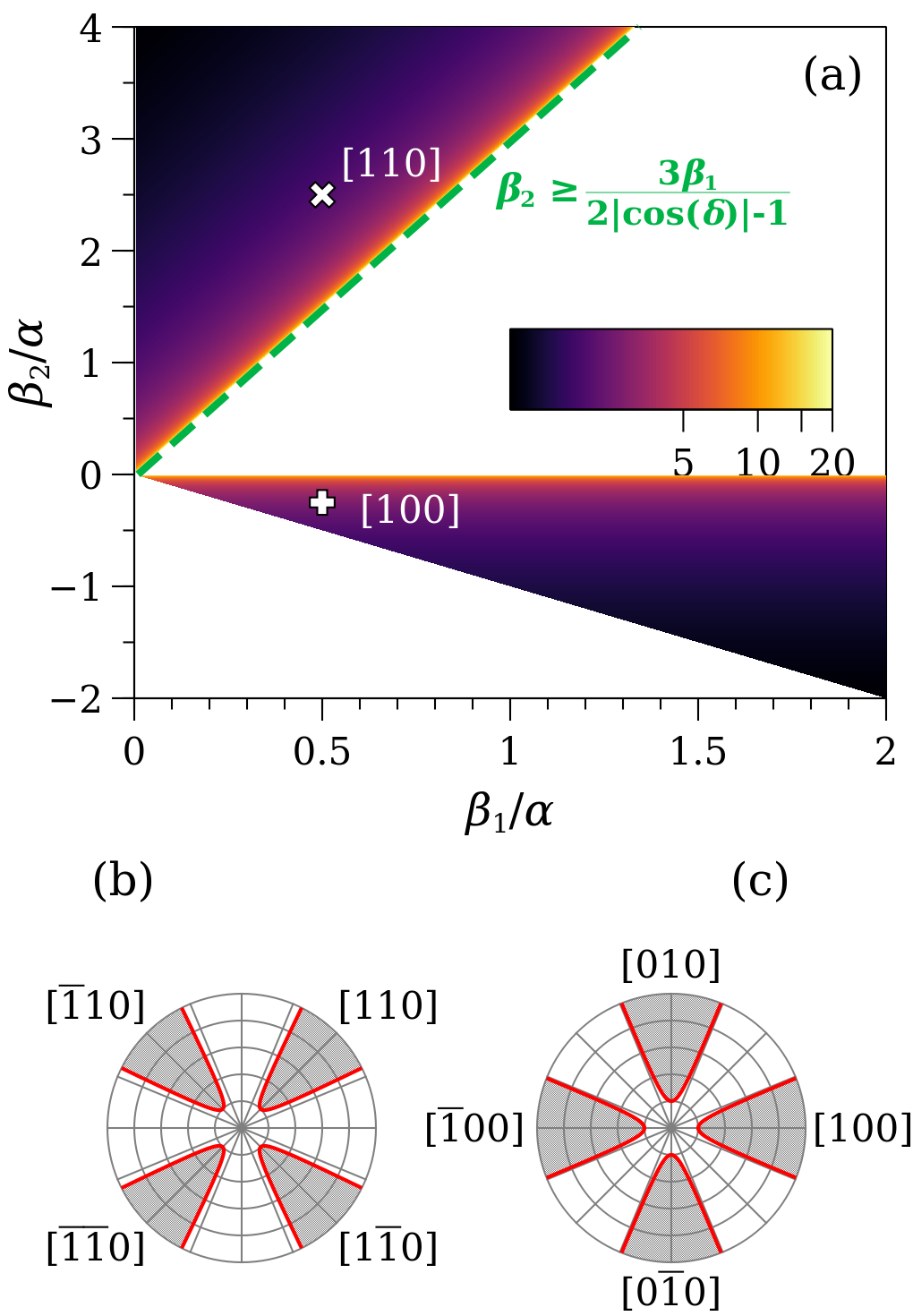}
    \caption{Dependence of critical electric field on GD parameters for $\delta = 0$. (a) Optimal magnitude of the dimensionless electric field $\bar{E} = \varepsilon_{0} \abs{\chi^{\mathrm{ret}}(\omega)} \, E_{0}/P_{0}$ (color of the intensity plot) in anharmonic-parameter space. The lower triangular region has the optimal field direction along $\Bqty{1, 0, 0}$, while the upper triangular region has the optimal field direction along $\Bqty{1, 1, 0}$. (b) Angular dependence of the critical field scaled by its value at the optimal angle for a representative point (white cross on Fig.~\ref{fig:CritElField}(a)), $\beta_{1}/\alpha = 0.5$, $\beta_{2}/\alpha = 2.5$. The shaded region depicts electric fields that stabilize the FE state. (c) Same as panel (b), for a representative point (white plus on Fig.~\ref{fig:CritElField}(a)),  $\beta_{1}/\alpha = 0.5$, $\beta_{2}/\alpha = -0.25$.}
    \label{fig:CritElField}
\end{figure}

\begin{figure*}[!ht]
    \centering
    \includegraphics[width=0.9\linewidth]{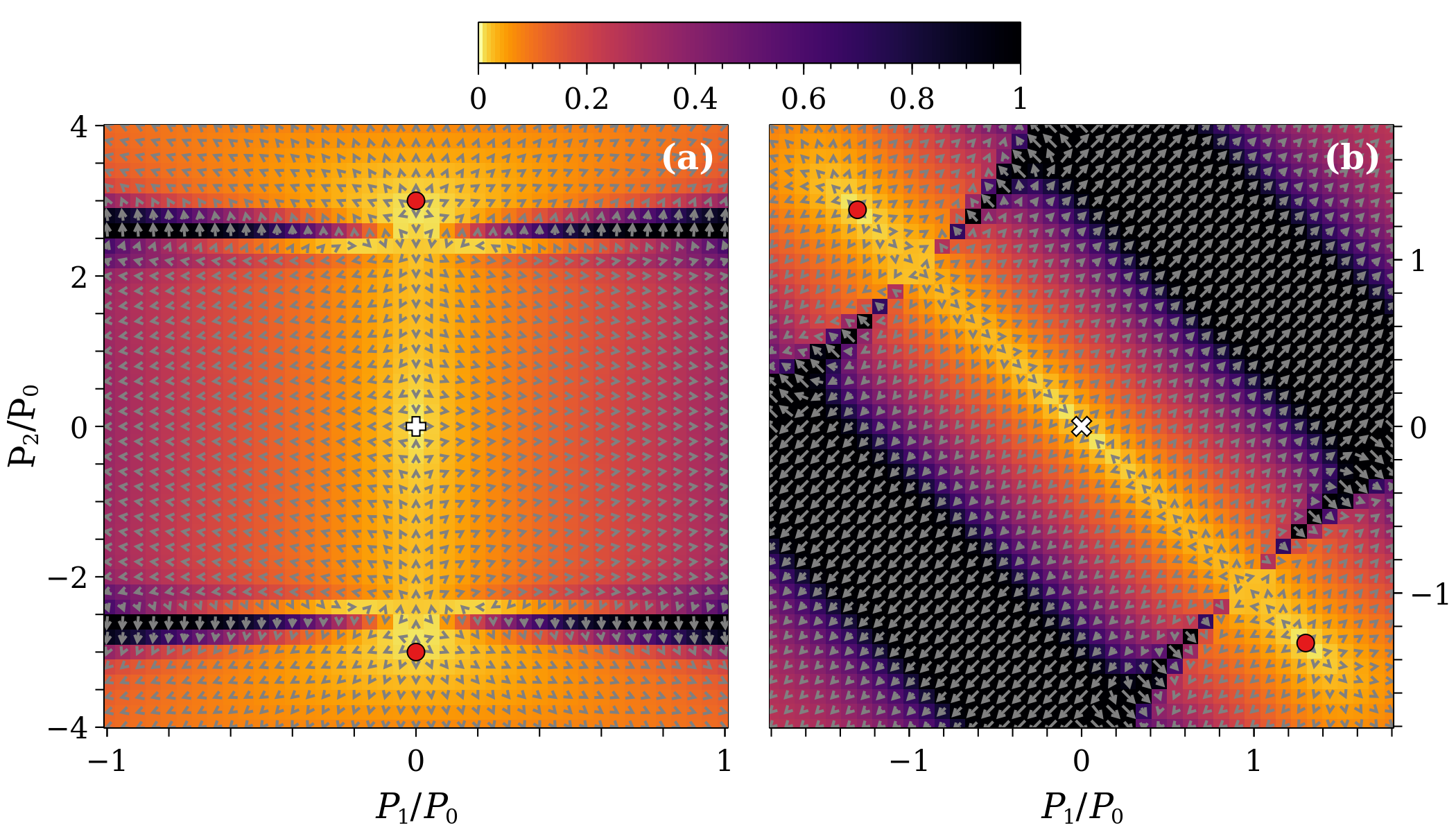}
    \caption{Plots of the vector field, colored by its magnitude, of the gradient of the effective free energy \req{eq:EffFreeEnGradient} under periodic drive for values of the parameters $\beta_{1,2}/\alpha$, together with the corresponding optimal electric field direction and critical electric field, corresponding to the white plus (a), and white cross (b), respectively, on the parameter diagram Fig.~\ref{fig:CritElField}(a). The magnitude of the drive is $E_{0}/E_{0, cr} = 3.0$, and the frequency of the drive is $\omega/\omega_{0} = 3.0$ with a small damping $r/\omega_{0} = 0.05$. We have checked that the choice of exact harmonic ratio of frequencies does not drastically alter the effect. The vector field is scaled according to the non-linear scaling $v_{i} = \mathrm{sgn}\qty(v_{i}) \, \qty(1 - e^{-\frac{\abs{v_{i}}}{v_{0}}})$, with a suitable scaling factor $v_{0}$. The white plus (cross) indicates the saddle point of the paraelectric phase, while the red points are the new stable fixed points.}    \label{fig:FreeEnGradient}
\end{figure*}
\reqs{eq:RenormSusc}, \rref{eq:Secondmoment}, \rref{eq:EffFreeEnGradient} are valid for any general linear susceptibility and anharmonic free energy density, and for arbitrary optical polarization $\vb{\pi}$. For a quartic anharmonic free energy, the third derivative in \req{eq:EffFreeEnGradient} is linear in $\expval{\vb{P}}$, and, thus, the Kapitza effect amounts to renormalization of the zero-Matsubara-frequency inverse susceptibility
\begin{subequations}
\label{eq:RenormZeroFSuscexpr}
\begin{equation}
    \qty[\chi^{M}_{\mathrm{eff}, 0}\qty(\expval{\vb{P}})]^{-1}_{i, j} = \frac{1}{\zeta_{i} \, \zeta_{j}} \, \Bqty{\qty(\chi^{M}_{0})^{-1}_{i, j} + \qty[\Pi\qty(\expval{\vb{\bar{P}}})]_{i, j}}, \label{eq:RenormZeroFSusc}
\end{equation}
where
\begin{eqnarray}
    & \Pi_{i, i} = 3 \beta_{1} \, \frac{\expval{\tilde{\bar{p}}_{i} \, \tilde{\bar{p}}_{i}}}{\zeta^{2}_{i}} + \beta_{2} \, \sum_{j \neq i} \frac{\expval{\tilde{\bar{p}}_{j} \, \tilde{\bar{p}}_{j}}}{\zeta^{2}_{j}}, \label{eq:DiagElemChiCorrect} \\
    & \Pi_{i, j} = 2 \beta_{2} \, \frac{\expval{\tilde{\bar{p}}_{i} \, \tilde{\bar{p}}_{j}}}{\zeta_{i} \, \zeta_{j}}, \ i \neq j. \label{eq:OffDiagElemChiCorrect}
\end{eqnarray}
We note that, for a non-zero $\expval{\vb{P}}$, the second moments \req{eq:Secondmoment} are functions of $\expval{\vb{P}}$, so the term $\qty[\chi^{M}_{0, \mathrm{eff}}\qty(\expval{\vb{P}})]^{-1}_{i, j} \, \expval{P}_{j}$ in the gradient of the free energy is still a highly non-linear function of $\expval{\vb{P}}$. The logic behind going to \reqs{eq:RenormZeroFSuscexpr} from \req{eq:EffFreeEnGradient}, and the detailed form of the third derivative are expounded in Appendix~\ref{App:SuscRenorm}.
\end{subequations}

\subsection{\label{subsec:PEInst}Instability of the paraelectric state}
We start by investigating the stability of the PE ($\expval{\vb{P}} = 0$) point, by considering the sign of the eigenvalues of \reqs{eq:RenormZeroFSuscexpr}, using \req{eq:Secondmoment}. Because of  \req{eq:RenormSusc}, the retarded susceptibility is not renormalized in the PE state.

For definiteness, the plane of optical polarization is chosen to be the $a-b$ plane. In that case, all the second moments $\expval{\tilde{\bar{p}}_{i} \, \tilde{\bar{p}}_{3}}$ are zero for $i = 1, 2, 3$. This decouples the $P_{3}$ mode from the $a-b$ plane modes. Furthermore, in the tetragonal phase $\zeta_{1} = \zeta_{2} = 1$. Then $\qty[\chi^{M}_{\mathrm{eff},0}]^{-1}_{3,3}$ in the $c$-direction is not renormalized. For the correction of the zero-Matsubara-frequency susceptibility in the $a-b$ plane, we have a $2\times2$ sub-matrix
\begin{subequations}
\label{eq:PESuscTransverse}
\begin{eqnarray}
    & \hat{\sigma}_{0} + \frac{\bar{E}^{2}}{2} \, \left[\frac{3 \beta_{1} + \beta_{2}}{2\alpha} \, \hat{\sigma}_{0} + \qty(\vec{d}_{\phi} \cdot \hat{\vec{\sigma}}) \right], \label{eq:PESuscMatForm} \\
    & \vec{d}_{\phi} = \langle \frac{\beta_{2}}{\alpha} \, \cos\qty(\delta) \, \sin\qty(2\phi), 0, \frac{3 \beta_{1} - \beta_{2}}{2\alpha} \, \cos\qty(2 \phi) \rangle, \label{eq:dVector}
\end{eqnarray}
\end{subequations}
where $\bar{E} = \qty(\varepsilon_{0} \, \abs{\chi^{\mathrm{ret}}\qty(\omega)} \, E_{0})/P_{0}$ is a dimensionless measure of the amplitude of the electric field of laser light. We also note here that incoherent light with randomized $\phi$ (and $\delta$) will lead to averaged matrix  \reqs{eq:RenormZeroFSuscexpr} having vanishing off-diagonal components, corresponding to a zero $\vec{d}_{\phi}$ in \req{eq:dVector}, and inhibiting the possibility of Kapitza stabilization.

The eigenvalue of the tensor \req{eq:PESuscTransverse} that can become negative has a polarization direction $\hat{\vb{p}}_{\mathrm{FE}} = \langle \cos\qty(\psi), \sin\qty(\psi), 0\rangle$, where
\begin{subequations}
\label{eq:PEInstability}
\begin{eqnarray}
    & \cos\qty(2\psi) = -\frac{\qty(d_{\phi})_{3}}{\abs{\vec{d_{\phi}}}}, \label{eq:InstDir1} \\
    & \sin\qty(2\psi) = -\frac{\qty(d_{\phi})_{1}}{\abs{\vec{d_{\phi}}}}, \label{eq:InstDir2}
\end{eqnarray}
and the condition for instability is
\begin{eqnarray}
    \frac{\bar{E}^{2}}{2} \, \qty[\abs{\vec{d}_{\phi}} - \frac{3 \beta_{1} + \beta_{2}}{2 \alpha}] \ge 1. \label{eq:InstCond}
\end{eqnarray}

% Discussion of optimal direction
The smallest magnitude $\bar{E}$ that fulfills the instability condition \req{eq:InstCond} is in the direction $\phi$ for which the term in the square brackets acquires the largest positive value. From \req{eq:dVector}, it is easy to see that the angular-dependent part of that expression is
\[
\abs{\vec{d}_{\phi}} = \sqrt{\qty(\frac{\beta_{2}}{\alpha} \, \cos\qty(\delta))^{2} \, \sin^{2}\qty(2 \phi) + \qty(\frac{3 \beta_{1} - \beta_{2}}{2\alpha})^{2} \, \cos^{2}\qty(2\phi)}
\]

Depending on whether $\frac{\abs{\beta_{2}}}{\alpha} \, \abs{\cos\qty(\delta)}$ or $\frac{\abs{3 \beta_{1} - \beta_{2}}}{2 \alpha}$ is greater, the optimal direction for the electric field is $\phi_{\mathrm{opt}} = \pi/4$ (direction $\qty[110]$), or $\phi_{\mathrm{opt}} = 0$ (direction $\qty[100]$).
% Discussion of region for orientation [110]
Then, \req{eq:InstCond} has two cases
\begin{eqnarray}
    & \frac{\qty(\bar{E}_{\qty[110]})^{2}}{2} \, \qty[\abs{\cos\qty(\delta)} \, \frac{\abs{\beta_{2}}}{\alpha} - \frac{3 \beta_{1} + \beta_{2}}{2 \alpha}] \ge 1, \label{eq:InstCond11} \\
    & \frac{\qty(\bar{E}_{\qty[100]})^{2}}{2} \, \qty[\frac{\abs{3 \beta_{1} - \beta_{2}}}{2\alpha} - \frac{3 \beta_{1} + \beta_{2}}{2 \alpha}] \ge 1. \label{eq:InstCond10}
\end{eqnarray}
The condition corresponding to \req{eq:InstCond11} is fulfilled when the term in the square brackets is positive. This delimits three regions for the parameter $\beta_{2}$: $\beta_{2} < -3\beta_{1}/(1 + 2 \abs{\cos\qty(\delta)})$, regardless of the value of $\cos\qty(\delta)$; $0 < \beta_{2} < 3 \beta_{1}/\qty(1 - 2 \abs{\cos\qty(\delta)})$ when $\abs{\cos\qty(\delta)} < 1/2$; and $\beta_{2} > 3\beta_{1}/\qty(2\abs{\cos\qty(\delta)} - 1)$ when $\abs{\cos\qty(\delta)} > 1/2$. The case for negative $\beta_{2}$ is unfeasible on grounds that the quartic GD free energy \req{eq:QuarticFEDens} for the tetragonal phase is not bounded from below in the polarization direction $\qty[110]$ for $\beta_{2} < -\beta_{1}$ (which explains the lower diagonal limit in Fig.~\ref{fig:CritElField}(a)). We still have to verify that $\abs{\beta_{2}} \, \abs{\cos\qty(\delta)} > \abs{3 \beta_{1} - \beta_{2}}/2$ when $\beta_{2} > 0$. This limits $\beta_{2}$ to $0 < 3\beta_{1}/\qty(1 + 2 \abs{\cos\qty(\delta)}) < \beta_{2} < 3 \beta_{1}/\qty(1 - 2 \abs{\cos\qty(\delta)})$ when $\abs{\cos\qty(\delta)} < 1/2$, and to $\beta_{2} > 3 \beta_{1}/\qty(2 \abs{\cos\qty(\delta)} + 1)$ when $\abs{\cos\qty(\delta)} > 1/2$. We see that the constraints when $\abs{\cos\qty(\delta)} < 1/2$ are contradictory. Conversely, the conditions in the case $\abs{\cos\qty(\delta)} > 1/2$ reduce to the most stringent condition $\beta_{2} > 3\beta_{1}/\qty(2 \abs{\cos\qty(\delta)} - 1)$, which is depicted by the green dashed line in Fig.~\ref{fig:CritElField}(a).

The first region is delimited by a dashed green line in Fig.~\ref{fig:CritElField}.

The magnitude of the critical electric field is obtained by the value that saturates the inequality \req{eq:InstCond11}
\begin{eqnarray}
& \qty(\bar{E}_{\qty[110]})_{\mathrm{cr}} = \sqrt{\frac{4 \alpha}{\beta_{2} \, \qty(\abs{2 \cos\qty(\delta)} - 1) - 3 \beta_{1}}},  \label{eq:Ecrit11} \\
& \abs{\cos\qty(\delta)} > \frac{1}{2}. \nonumber
\end{eqnarray}
From \reqs{eq:InstDir1}, \rref{eq:InstDir2}, we see that the instability polarization direction is $\psi = -\pi/4$ (direction $\qty[\bar{1}10]$), which is perpendicular to the direction of the applied electric field vector.

% Discussion of region for orientation [100]
A similar analysis can be performed for the condition corresponding to \req{eq:InstCond10}. The expression in the square brackets is positive when $\beta_{2} < 0$. We still have to verify that $\abs{\beta_{2} - 3 \beta_{1}}/2 > \abs{\beta_{2}} \, \abs{\cos\qty(\delta)}$ when $\beta_{2} < 0$. This does not pose any more stringent constraints on $\beta_{2}$ for $\abs{\cos\qty(\delta)} < 1/2$, but it poses $\beta_{2} < - 3\beta_{1}/\qty(2\abs{\cos\qty(\delta)} - 1) < -3\beta_{1}$, which is outside the region of thermodynamic stability. The magnitude of the critical electric field is then
\begin{eqnarray}
& \qty(\bar{E}_{\qty[100]})_{\mathrm{cr}} = \sqrt{\frac{2 \alpha}{-\beta_{2}}}, \label{eq:Ecrit10} \\
& \abs{\cos\qty(\delta)} < \frac{1}{2}. \nonumber
\end{eqnarray}
and the instability polarization direction is $\psi = \pi/2$ (direction $\qty[010]$), again, perpendicular to the applied electric field vector.

\end{subequations}
The intensity plot of these optimal fields is plotted in Fig.~\ref{fig:CritElField}(a).

For directions of the applied field other than the optimal direction, the critical electric field magnitude is larger than the optimal, and Figs.~\ref{fig:CritElField}(b, c) depict the angular dependence of the magnitude of the critical electric field to the optimal one (red lines) for a representative point of the upper (lower) triangular regions in Fig.~\ref{fig:CritElField}(a) (white cross, and white plus, respectively). We see that there are asymptotic bounds around the optimal directions, and the grey regions of the polar plot indicate combinations of direction and magnitude of the applied electric field that ought to stabilize an FE state.

We note that for sufficiently small magnitudes of the cross-coupling term $\beta_{2}$ in \req{eq:QuarticFEDens} (the excluded white region in Fig.~\ref{fig:CritElField}(a) between the two allowed triangular regions), we cannot have Kapitza stabilization.

\subsection{\label{subsec:FEStable}Ferroelectric fixed points}
Fixing the electric field beyond its critical value, we focus on the gradient field of the effective free energy \req{eq:EffFreeEnGradient}.

For the same choice of parameters labeled in Fig.~\ref{fig:CritElField} by the white plus for the $\qty[100]$ direction and by the white cross in the $\qty[110]$ direction, the corresponding gradient field is depicted in Fig.~\ref{fig:FreeEnGradient}. As predicted by the analysis in Sec.~\ref{subsec:PEInst}, \reqs{eq:InstDir1}-\rref{eq:InstDir2}, the instability direction is in the $a-b$ plane in a perpendicular direction to the applied electric field. We note that in both plots the PE point ($\vb{P} = \vb{0}$) becomes a saddle-point, because in the direction of the applied electric field, there is only mode stiffening. This behavior of the gradient of the effective free energy is quite different than for the free energy in the absence of driving fields, which further reinforces the finding that cross-coupling of different polarization directions is crucial for Kapitza stabilization within this model. Additionally, the stable (saddle) points of the GD free energy, $\sqrt{-\alpha/\beta_{1}} \, \qty[100]$, or $\sqrt{-\alpha/\qty(\beta_{1} + \beta_{2})} \, \qty[110]$, are no longer extremal points of the effective free energy density.

\subsection{\label{subsec:DFTcalc}Estimation of GD parameters from ab-initio calculations}
We estimate the coefficients in the GD free energy density expression for the case of STO. The harmonic part of the GD free energy is given by the $q = 0$ contribution in \req{eq:HarmAction}, remembering that $\Gamma_{0} = \qty(L^{3}/T) \, \mathcal{F}_{0}$, and $\vb{P}_{q = 0} = \sqrt{L^{3}/T} \, \vb{P}$. The anharmonic part takes the form of  \req{eq:QuarticFEDens}. The fitting model then has the form
\begin{subequations}
\label{eq:FittingFunc}
\begin{eqnarray}
    & \mathcal{F}_{\mathrm{GD}} = \frac{\alpha'}{2} \, \sum_{i} P^{2}_{i} \nonumber \\
    & + \frac{\beta'_{1}}{4} \, \sum_{i} P^{4}_{i} + \frac{\beta'_{2}}{2} \, \sum_{i < j} P^{2}_{i} \, P^{2}_{j}, \label{eq:GDFreeEnUC} \\
    & \alpha' = \frac{\alpha}{\varepsilon_{0}}, \label{eq:SquareTerm} \\
    & \beta'_{i} = \frac{\beta_{i}}{\varepsilon_{0} \, P^{2}_{0}}, \ \qty(i = 1, 2). \label{eq:QuarticTerms}
\end{eqnarray}
\end{subequations}

We begin by using DFT to relax the internal coordinates of cubic and tetragonal (including antiferrodistortive rotations of the oxygen octahedra) unit cells, both in their centrosymmetric reference structures as well as with a displacement of the Ti atoms added in the [100] or [110] directions. The displacements break the symmetry and lead to a polarization in the respective directions. The size of the polarization is calculated based on the atomic displacements and fixed Born effective charges.
The free-energy density as a function of polarization is then obtained by interpolating between the corresponding centrosymmetric and polar structures, and is shown in Fig.~\ref{fig:sscha}. The details of the DFT calculation are given in Appendix~\ref{App:DFT}.

One challenge is that standard DFT does not include temperature or quantum ion effects, and so predicts STO to be ferroelectric, yielding a negative quadratic term \req{eq:SquareTerm}. To capture the experimental quantum paraelectric behavior, and a positive quadratic term \req{eq:SquareTerm}, we use the Stochastic Self-Consistent Harmonic Approximation (SSCHA) in combination with machine learning force fields to include these effects in an ab-initio description of STO at a reasonable computational cost~\cite{CesareSTO}. The details of these calculations are explained in Appendices~\ref{App:SSCHA} and \ref{App:ML Force Field}.   

To obtain a corresponding energy versus polarization curve within the SSCHA method, for which the structure relaxes as expected to the non-polar phase, we calculate the SSCHA energies for the DFT-calculated polar/centrosymmetric structures scaled to the experimental lattice parameters~\cite{SrTiO3_icsd}. The results are shown in Fig.~\ref{fig:sscha}.
% TABLE GD COEFFICIENTS -- BEGIN
\begin{table}[!ht]
\centering
\begin{ruledtabular}
\begin{tabular}{l c S[table-format=1.1,table-column-width=0.17\columnwidth] S[table-format=1.1,table-column-width=0.17\columnwidth] S[table-format=1.2,table-column-width=0.17\columnwidth]}
 & & {$\alpha' \times 10^{-8}$} & {$\beta'_{1} \times 10^{-10}$} & {$\beta'_{2} \times 10^{-10}$} \\
Structure & Method & {[$\unitAlpha$]} & {[$\unitBeta$]} & {[$\unitBeta$]} \\
\hline
Cubic & DFT  & -1.0 & 1.6 & 0.47 \\
Tetragonal & DFT & -1.7 & 1.9 & 0.60 \\
Tetragonal & SSCHA & 2.8 & 1.4 & 0.35 \\
\end{tabular}
\end{ruledtabular}
\caption{Fit parameters of the model \reqs{eq:FittingFunc} resulting from the curves shown in Fig.~\ref{fig:sscha}.}
\label{tab:fits}
\end{table}
% TABLE GD COEFFICIENTS -- END

\begin{figure}[t!]
    \centering
    \includegraphics[width=0.99\columnwidth]{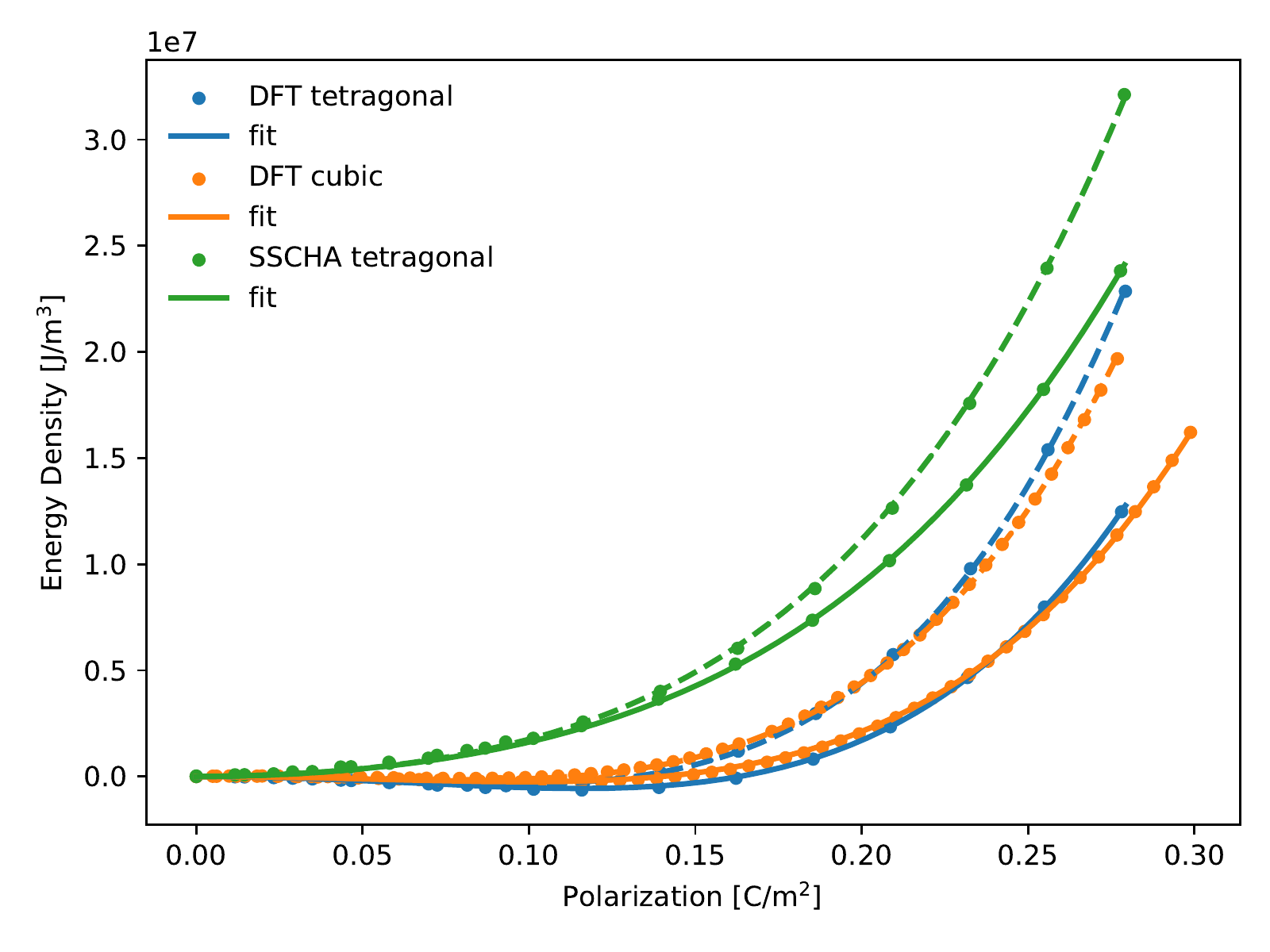}
    \caption{Energy versus polarization curves for the [110] (continuous line) and [100] (dashed line) directions calculated with DFT for the tetragonal (blue) and cubic (orange) phase and with SSCHA (green) for the tetragonal phase. The calculated results are presented as dots, while the fits to the data, which yielded the parameters listed in Table~\ref{tab:fits}, are shown as lines. The energies are plotted relative to the energy of the nonpolar structure.}
    \label{fig:sscha}
\end{figure}

We present the fit parameters in Table~\ref{tab:fits}. As expected, the quadratic term in the DFT case is negative indicating a characteristic double well potential with a ferroelectric instability from the non-polar phase. SSCHA on the other hand successfully yields a positive quadratic term. It also provides values for $\beta'_2$ and $\beta'_1$ which are both positive, but in a ratio of considerably less than the three required (Fig.~\ref{fig:CritElField}(a)) for the Kapitza effect. Searches for other materials must therefore be undertaken. Interestingly  Interestingly, the quartic coefficients obtained with DFT and SSCHA are similar, with $\beta_1$ almost identical, and  $\beta_2$ reduced by $\sim 30$\% in SSCHA compared with DFT. We therefore expect that scans for further suitable materials with appropriate $\beta$ values can be performed at the DFT level, with SSCHA used to check the values for the most promising candidates.

\subsection{\label{subsec:QCPcontrol}Control of the quantum critical point with light}
%New section -- discussion on QCP control
\begin{figure}[!htp]
    \centering
    \includegraphics[width=0.9\columnwidth]{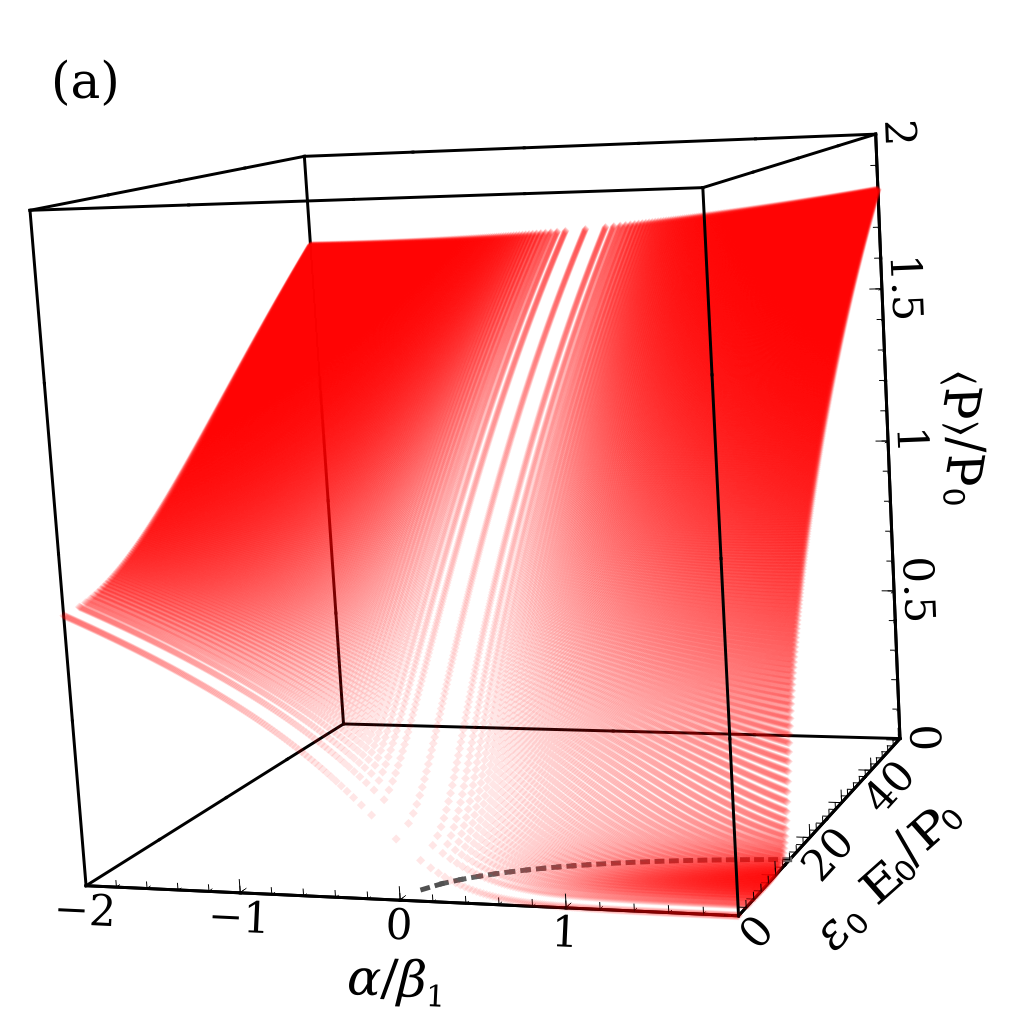} \\
    \includegraphics[width=0.9\columnwidth]{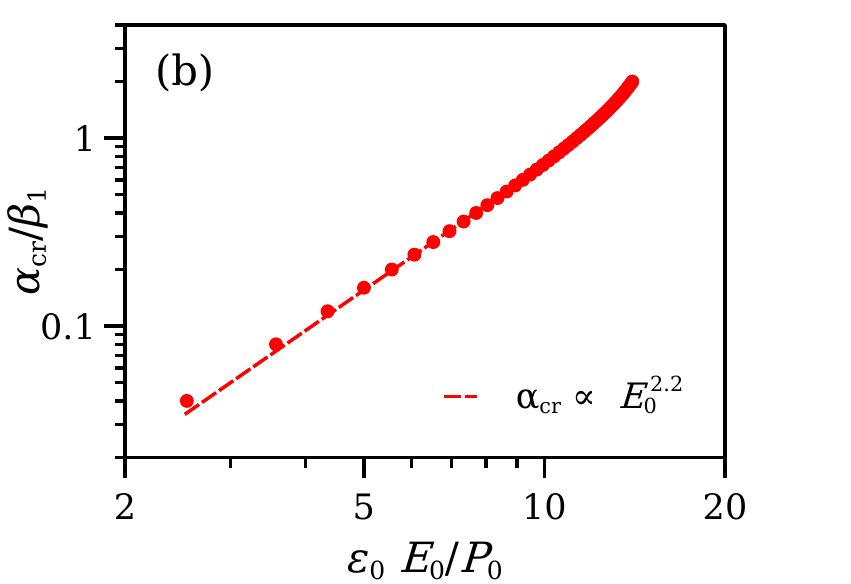} \\
    \includegraphics[width=0.9\columnwidth]{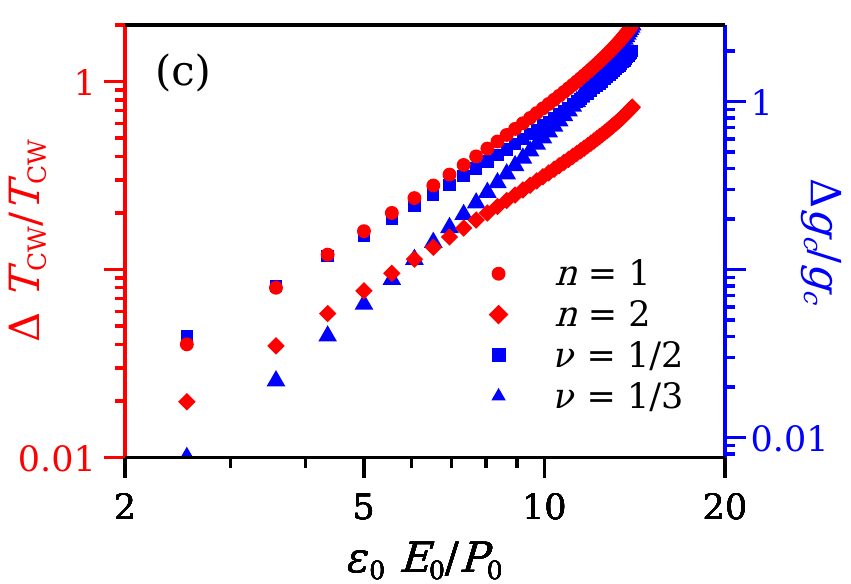}
    \caption{\label{fig:QCPcontrol}Control of PE-FE transition with applied light. We choose a ratio of the quartic coefficients $\beta_{2}/\beta_{1} = 5.0$ with the optimal polarization direction of the electric field vector in the $\qty[\bar{1}10]$ and linearly polarized light ($\delta = 0$). (a) Stable FE order $\expval{P}$ as a function of the magnitude of light's electric field $E_{0}$ and quadratic coefficient $\alpha$. The dashed line in the $\expval{P} = 0$ plane indicates a critical line. (b) The critical line from panel (a) presented on a log-log plot, with a linear fit indicated by a dashed line with the corresponding power. (c) Scaling of the critical temperature (left vertical axis (red)), or critical control parameter $g_{c}$ (right vertical axis (blue)) for several values of $n$ and $\nu$ in \req{eq:CritExpScalingAlpha}.}
\end{figure}
The homogeneous, zero-Matsubara frequency inverse susceptibility $\alpha$ has characteristic scaling with temperature $T$ for finite temperature or a control parameter $g$ at zero temperature. Whenever it changes sign signals a critical point, either thermal or quantum-critical. Conversely, one can utilize $\alpha$ as a scaling variable to model the temperature or control parameter. Therefore, a helpful way to visualize the effect of the applied electric field is to plot the stable fixed point $\expval{P}/P_{0}$ as a function of $\alpha/\beta_{1}$ and electric field magnitude $\varepsilon_{0} \, E_{0}/P_{0}$ for a particular choice of the ratio of quartic coefficients, as is done in Fig.~\ref{fig:QCPcontrol}(a). For negative $\alpha$, we see that applying an electric field enhances the value of the FE order. For positive $\alpha$ (PE state), there is a critical region in the $\alpha$-$E$ plane where the stable order is zero. The line is simply the condition that the smaller eigenvalue of \req{eq:PESuscMatForm} becomes zero, and we depict the power-law scaling of this curve in Fig.~\ref{fig:QCPcontrol}(b). The deviation of the exponent from $2$ is due to the dependence of the susceptibility on $\alpha$. For small values of $\alpha$, it is well approximated by $2$.
	
We attribute the shift of the critical value for $\alpha$ with $E$ as a shift of the critical transition point induced by light irradiation. This suggests a convenient parametrization of the dependence of the critical field $E = E_{\mathrm{cr}}\qty(\tilde{\alpha})$ on a control parameter $T$ or $g$ as a function of $\tilde{\alpha}$ in \reqs{eq:Ecrit11}, \rref{eq:Ecrit10}, and inverting by inverting \req{eq:CritExpScalingAlpha} with respect to $T$ or $g$ as a function of the parameter %\tilde{\alpha}$. This can be due to
\begin{subequations}
\label{eq:QCPScalings}
\begin{enumerate}
    \item{
    Thermal transition at finite temperature and for the control parameter $g < g_{c}$. From \req{eq:CritExpScalingAlpha}, we can read off (note that $\alpha\qty(0, g) < 0$ for $g < g_{c}$)
    \begin{equation}
        \frac{\Delta T_{\mathrm{CW}}\qty(g, \tilde{\alpha})}{T_{CW}\qty(g, 0)} = \qty[1 - \frac{\tilde{\alpha}}{\alpha\qty(0, g)}]^{\frac{1}{n}} - 1. \label{eq:TcScaling}
    \end{equation}
    }
    \item{Quantum phase transition at zero temperature and for a control parameter $g > g_{c}$. Again, from \req{eq:CritExpScalingAlpha}, we can read off
    \begin{equation}
        \frac{\Delta g_{c}\qty(\tilde{\alpha})}{g_{c}\qty(0)} \propto \tilde{\alpha}^{\frac{1}{2\nu}}. \label{eq:gcScaling}
    \end{equation}
    }
\end{enumerate}
\end{subequations}
Both of these scalings are depicted in Fig.~\ref{fig:QCPcontrol}(c). For small values of $\tilde{\alpha}$, the power-law dependence of the critical temperature is $2$, regardless of the values of the non-universal exponent $n$, whereas the power-law dependence of the critical value of the control parameter $1/\nu$.

\subsection{\label{subsec:SecondHarmonic}Experimental signatures of the polarized state}

Concerning experimental observations of the proposed effect, we suggest the following experimental tests: (a) conventional dielectric measurements, (b) optical second harmonic generation, and (c) X-ray diffraction. 

a) In the case of an electrical measurement, one performs a poling measurement of the accumulated charge across a capacitor in an unbiased setup. The charge saturates during the irradiation and should vanish when irradiation is interrupted. The displacement current, which is the derivative of the charge, should show oppositely directed peak signals during the start and end of the irradiation. A sketch of the proposed setup and expected time traces of the signal are presented in Appendix~\ref{App:PolingSetup}.

 b) In this case, a second harmonic to a weak signal at a frequency different than the drive frequency is generated when the sample is FE, with signal strength proportional to the induced polarization and specific angular dependence. Using the quantum effective action \req{eq:EffAction}, we have obtained an expression for the non-linear susceptibility tensor, and the details are discussed in Appendix~\ref{App:SecondHarmonic}.
 
%NEW PARAGRAPH
(c) X-ray diffraction~\cite{McWhanJPC1985} measures directly the lattice strain associated with FE polarization and is nowadays routinely performed at free electron lasers with femtosecond time resolution. The experiment here will involve pumping the sample with THz radiation and then measuring the position of the principal Bragg peaks as a function of the time between the pump pulse and the X-ray pulse. Because of the tensor character of the strain, which is mirrored in the resulting displacements of different Bragg peaks, it will be possible to reconstruct the full strain tensor as a function of experimental parameters such as THz pulse length and amplitude.

The strong coupling between strain and electric polarization is an important feature of the titanates and other (near) ferroelectrics. Generically, the latter builds up first, followed by the appearance of strain~\cite{McWhanJPC1985, AbernathyEPL_1988}, which in turn stabilizes the polarization~\cite{NovaSci2019,LiSci2019,XuNatComm2020}. This has two benefits: the first is that there is no need for a continuous power supply to maintain the FE state, and the second is that there is ample time to probe whether a FE state has actually been realized experimentally.

%Electric field intensity
As far as an order-of-magnitude estimate for the necessary electric field magnitude, we use the following estimates for the GD free energy parameters~\cite[see Table I in Supplementary Information]{RowleyNatPhys2014}, $a = \alpha = 5.1 \times 10^{-5}$, $b = \beta_{1}/P^{2}_{0}$. Taking $\beta_{1}/\alpha = 0.5$, implies $P_{0} = \sqrt{\qty(\beta_{1}/\alpha) \, a/b} = 1.9 \, \mu{\mathrm{C}}/\mathrm{cm}^{2}$. Then, using a magnitude of a complex permittivity $\tilde{\varepsilon}_{\omega} = -60 + i \, 40$ at a drive frequency of around $\omega/2\pi = 3.8 \, \mathrm{THz}$, which is twice the frequency of the soft-mode phonon~\cite{FedorovFE1998}, and a dimensionless critical field of the order of  $\bar{E}_{cr} = 2.8$ (see Fig.~\ref{fig:CritElField}), we get an absolute electric field magnitude $E_{0, \mathrm{cr}} = \bar{E}_{\mathrm{cr}} \, P_{0}/\qty(\varepsilon_{0} \, \abs{\chi^{\mathrm{ret}}\qty(\omega)})  = 830 \, \mathrm{kV}/\mathrm{cm}$.

%Heating power density
For the estimate of the heat deposited in the sample by the laser field we use the Joule heating power density $q = \sigma_{\omega} \, E^{2}_{0}/2$, where $\sigma_{\omega} = \omega \, \varepsilon_{0} \, \varepsilon''_{\omega}$ is the electric conductivity of the sample, which is estimated to be around $8.5 \times 10^{3} \, \mathrm{\Omega}^{-1} \, \mathrm{m}^{-1}$, giving an estimate for $q = 29 \, \mathrm{W}/\mathrm{\mu{m}}^{3}$. To get an estimate of the heat fluence $\phi_{q}$, one has to remember that the electric field attenuates in the sample with a characteristic length scale $\frac{1}{\lambda_{\omega}} = \frac{\omega}{c} \, \Im\qty[\sqrt{\tilde{\varepsilon}_{\omega}}]$, $\lambda_{\omega} = 0.18 \, \mu\mathrm{m}$. The integrated power density over the sample thickness $t$ and pulse duration $\tau$ gives $\phi_{q} = 2 \, \lambda_{\omega} \, \tau \, \qty(1 - e^{-\frac{t}{2 \lambda_{\omega}}}) \, q_{0}$. Assuming sample thickness much larger than $2\lambda_{\omega}$, and pulse duration of at least $10$ periods of the light field, the estimate for the fluence is $\phi_{q} = 2.7 \, \mathrm{mJ}/\mathrm{cm}^{2}$.

%NEW PARAGRAPH 
It is important to acknowledge that the necessary power levels per unit volume may not be practically feasible in continuous-wave set-up. As previously mentioned, we must rely on coupling to strain to maintain polarization after the Kapitza "pump" has been removed. Notably, experience with dynamically induced nucleation and growth (at resonance) suggests that the relevant time scales for growth and decay lie within the femtosecond to picosecond range. Consequently, terahertz radiation pulses lasting around 1 picosecond and resulting in a net energy dissipation of 3 picowatts per cubic micrometer should be sufficient to observe the predicted effects.

\section{\label{sec:Conclude}Conclusion}
In this study, we extend the ideas of Kapitza engineering to the effective field theory near a quantum critical point (QCP) to show how the incipient order can be induced by the off resonance  drive.  As a specific example we considered the case of the strong off-resonance photon field applied to  stabilize the ferroelectric (FE) phase in an incipient displacive FE.  We find that it could be feasible to induce the FE phase if the material is close to a QCP and has a sufficiently large cross-coupling term in the anharmonic portion of the free energy.

We predict that the critical light field will exhibit a sensitive dependence on the direction and phase of the polarization axis. When field intensities surpass the critical field, the polarization will develop along an axis in the plane of the light's polarization and perpendicular to the applied electric field. 

To make contact with materials we have developed detailed ab-initio microscopic models that enable us evaluate the realistic parameters of the Ginzburg-Devonshire action, such as $\beta_1,\beta_2$ coefficients, and force fields in
$\sto$. The advantage of the proposed approach is in combining realistic materials parameters with the effective GD modeling of Kapitza engineering of FE state. 
While Kapitza engineering in $\sto$ would be a challenge based on the estimated  parameters, we stress that we present the general framework. Our results: 

\begin{enumerate}
    \item[a)]{Establish a clear-cut numerical procedure that takes into account quantum-mechanical zero-point motion effects that render STO a quantum paraelectric;
    }
    \item[b)]{Demonstrate a deviation from isotropy (which corresponds to $\beta_{2} = \beta_{1}$) and are not far off from the required stability criterion. The parameters are of the same order on the ``wrong side" of the inequality Fig.~\ref{fig:CritElField}(a), yet they are close. We draw the conclusion that the working regime in a similar class of quantum paraelectrics is possible. It is also possible that the choice of the functional would affect the relation between $\beta_{2},  \beta_{1}$) and would bring them in the desired range, where Kapitza stabilization would be possible;
    }
    \item[c)]{These results together with the ab initio work serve as a guide for a well-defined process to be followed in different perovskite materials and under varying constraints in the search for suitable materials and optimal experimental constraints to make Kapitza stabilization possible.}
\end{enumerate}

We propose experimental signatures, such as poling measurements of accumulated polarization charge in an unbiased capacitor, second-harmonic generation in response to a weak probe pulse with an appropriate frequency and polarization axis, and X-ray diffraction measurements of the resulting strain. We also discuss the role of the Joule heating that would make the proposed set up impractical for the continuous-wave conditions. On the other hand, we estimate that finite-duration pulses with slow-strain dynamics could facilitate the stabilization of these phases. 

Our proposed approach can be extended to other quantum-critical orders, including magnetism and superconductivity. We anticipate that factors like linear vs. quadratic coupling to drive fields will depend on the order's symmetry. The necessary resonant field strengths can be achieved with current experimental capabilities, making this mechanism a complementary method for manipulating quantum matter using terahertz light. \emph{Note:} While this manuscript was in review, we became aware of a preprint~\cite{ZhuangarXiv2023} that is related to our work.
\begin{acknowledgments}
The authors benefited greatly from discussions with M.~Geilhufe, I.~Khaimovich, A.~Fisher, S.~Bonetti, M.~Basini, V.~Unikandanunni, A.~Cavallieri, E.~Demler, D.~Jaksch,  P.~B.~Littlewood, M.~Mitrano, P.~Moll, A.~Polkovnikov and P.~Volkov. This work was supported by the and by the European Research Council (ERC) under the European Union’s Horizon 2020 research and innovation program project HERO Grant Agreement No. 810451. DK and AB also acknowledge support by  VILLUM FONDEN via the Centre of Excellence for Dirac Materials (Grant No. 11744), University of Connecticut and the Swedish Research Council (VR) 2017-03997. JS was also supported by the Joachim Herz Foundation. Computational resources were provided by ETH Zurich and by the Swiss National Supercomputing Center (CSCS) under project ID s1128. We thank the authors of Ref.~\cite{CesareSTO} for providing their training and testing data.
\end{acknowledgments}
\section*{Data availability statement}
The supporting DFT, force field and SSCHA calculations for this article are openly available in the materials cloud entry \cite{materialscloud_entry}.
\bibliography{FE_QCP}

\appendix
\begin{figure}[h]
    \centering
    \includegraphics[width=0.8\columnwidth]{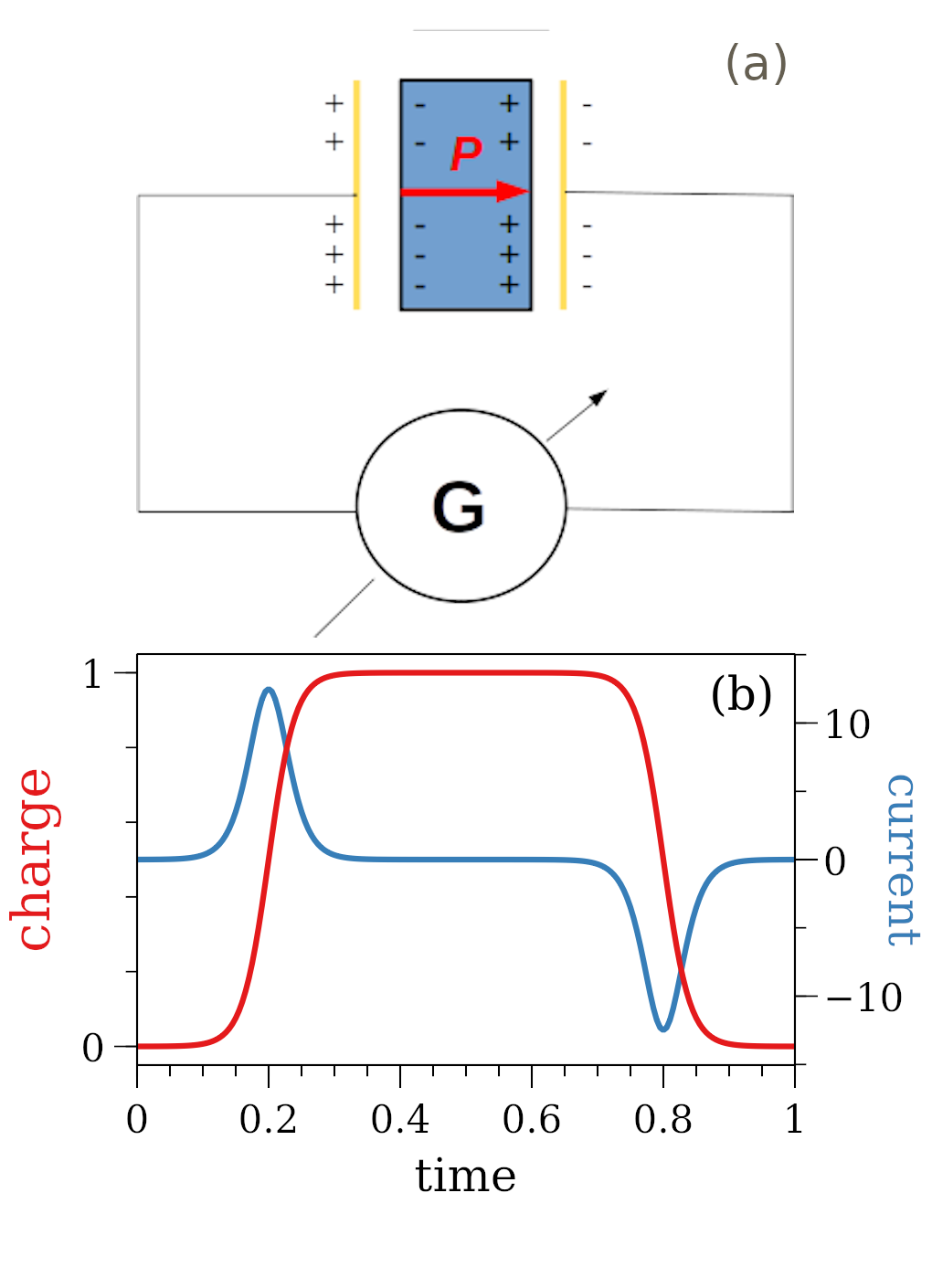}
    \caption{\label{fig:Poling}(a) Sketch of the unbiased poling measurement with a galvanometer (G) of an irradiated FE (blue sample) in a capacitative setup (gold electrodes). Illustrations of time traces for the accumulated charge and displacement current during irradiation}
\end{figure}
\section{\label{App:SuscRenorm}Renormalization of the zero-frequency susceptibility}
Going from \req{eq:EffFreeEnGradient} to \reqs{eq:RenormZeroFSuscexpr} relies on the fact that the anharmonic free energy is a quartic function of polarization $\expval{\vb{P}}$, and, so, its third derivative that enters in the third term in \req{eq:EffFreeEnGradient} is a linear function of $\expval{\vb{P}}$ that we can ascribe to a renoramlization of the coefficient in terms of the linear term, which is the first term in \req{eq:EffFreeEnGradient}. Having in mind the specific form of \req{eq:QuarticFEDens}, we can write the third derivatives in the following form
\begin{subequations}
\begin{eqnarray}
    & \frac{\partial^{3}}{\partial \bar{P}_{1} \, \partial \bar{P}_{j} \, \partial \bar{P}_{k}} \qty(\frac{\varepsilon_{0} \, \mathcal{F}_{\mathrm{ah}}}{P^{2}_{0}}) = \frac{2}{\zeta_{i} \, \zeta_{j} \, \zeta_{k}} \, \left\lbrace 3 \beta_{1} \, \frac{\bar{P}_{i}}{\zeta_{i}} \, \delta_{j, i} \, \delta_{k, i} \right. \nonumber \\
    & + \beta_{2} \, \left[ \frac{\bar{P}_{k}}{\zeta_{k}} \, \delta_{j, i} \, \qty(1 - \delta_{k, i}) + \frac{\bar{P}_{j}}{\zeta_{j}} \, \qty(1 - \delta_{j, i}) \, \delta_{k, i} \right. \nonumber \\
    & \left. \left.  + \frac{\bar{P}_{i}}{\zeta_{i}} \, \qty(1 - \delta_{i, j}) \, \delta_{j, k} \right] \right\rbrace. \label{eq:ThirdDerivFreeEn}
\end{eqnarray}
This third derivative \req{eq:ThirdDerivFreeEn} gets contracted by $\expval{\tilde{\bar{p}}_{j} \, \tilde{\bar{p}}_{k}}$ in the second term on the right-hand side of \req{eq:EffFreeEnGradient}. We get
\begin{eqnarray}
    & \frac{1}{2} \sum_{j, k} \expval{\tilde{\bar{p}}_{j} \, \tilde{\bar{p}}_{k}} \, \frac{\partial^{3}}{\partial \bar{P}_{1} \, \partial \bar{P}_{j} \, \partial \bar{P}_{k}} \qty(\frac{\varepsilon_{0} \, \mathcal{F}_{\mathrm{ah}}}{P^{2}_{0}}) \nonumber \\
    & = \frac{1}{\zeta_{i}} \, \left\lbrace 3 \beta_{1} \, \frac{\expval{\tilde{\bar{p}}_{i} \, \tilde{\bar{p}}_{i}}}{\zeta^{2}_{i}} \, \frac{\bar{P}_{i}}{\zeta_{i}} + \right. \nonumber \\
    & \left. + \beta_{2} \, \left[ \sum_{j \neq i} \frac{\expval{\tilde{\bar{p}}_{j} \, \tilde{\bar{p}}_{j}}}{\zeta^{2}_{j}} \frac{\bar{P}_{i}}{\zeta_{i}} + 2 \, \sum_{j \neq i} \frac{\expval{\tilde{\bar{p}}_{i} \, \tilde{\bar{p}}_{j}}}{\zeta_{i} \, \zeta_{j}} \, \frac{\bar{P}_{j}}{\zeta_{j}}\right]\right\rbrace \nonumber \\
    & = \frac{1}{\zeta_{i}} \, \sum_{j} \Pi_{i, j} \, \frac{\bar{P}_{j}}{\zeta_{j}}, \label{eq:KapitzaTermSimplify}
\end{eqnarray}
where we used a shorthand notation $\Pi_{i, j}$ for every coefficient in front of a $\bar{P}_{j}/\zeta_{j}$ term. These are different for $i = j$, and for $i \neq j$, and their form is given in \reqs{eq:DiagElemChiCorrect}, \rref{eq:OffDiagElemChiCorrect}
\end{subequations}

\section{\label{App:PolingSetup}Diagram of the poling setup}
For completeness, we illustrate the electrical measurement for detecting polarization. The schematic setup is presented in Fig.~\ref{fig:Poling}(a).

The expected measurement involves the measurement of accumulated charge on the capacitor plates when the FE develops a polarization $P$. The displacement current should show oppositely directed peaks at the turning on and turning off of the induced polarization, as schematically presented in Fig.~\ref{fig:Poling}(b).

\section{\label{App:SecondHarmonic}Second-harmonic-generation calculation details}
\begin{figure}[!ht]
    \centering
    \includegraphics[width=0.85\columnwidth]{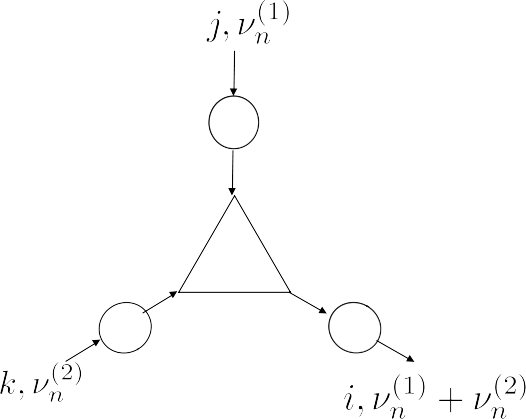}
    \caption{A Feynman diagram for the expression \req{eq:NonLinSuscFourier}, assuming $\vb{q}_{1} = \vb{q}_{2} = \vb{0}$ on the external labels. The triangle is the third derivative of the anharmonic free energy and is frequency-independent. The circles are linear susceptibilities at the corresponding frequency.}
    \label{fig:NonLinSusc}
\end{figure}
Here we present the theory for calculating the non-linear susceptibility tensor $\chi^{(2)}$ in the presence of a non-zero polarization $\expval{\vb{P}}$.We outline the steps to arrive at nonlineal susceptibility. We  leave specific features, that depend on experimental set up, for a separate discussion.

Susceptibility is a rank-three tensor that is defined as the third (functional) derivative of the cumulant-generating function $W\qty[\vb{E}]$
\begin{widetext}
\begin{equation}
    \chi^{\qty(2)}_{i, j, k}\qty(\vb{x}, \tau_{1}; \vb{x}_{2}, \tau_{2}; \vb{x}_{3}, \tau_{3}) = -\frac{\delta^{3} W\qty[\vb{E}]}{\delta E_{i}\qty(\vb{x}_{1}, \tau_{1}) \, \delta E_{j}\qty(\vb{x}_{2}, \tau_{2}) \, \delta E_{k}\qty(\vb{x}_{3}, \tau_{3})}. \label{App:NonLinSuscDef}
\end{equation}
Keeping in mind the Legendre transform relation
\begin{equation}
    -\Gamma\qty[\vb{P}] + \qty(\vb{P}, \vb{E}) = -W\qty[\vb{E}], \label{eq:LegendreTrans}
\end{equation}
where $\qty(\vb{P}, \vb{E})$ is a shorthand for a space-time integral $\int \dd{{}^{4}x} \vb{P}\qty(x) \cdot \vb{E}\qty(x)$, it is easily verified that the variational derivative of the effective action $\Gamma\qty[\vb{P}]$ satisfies the equation of motion
\begin{equation}
    \frac{\delta \Gamma\qty[\vb{P}]}{\delta P_{i}\qty(\vb{x}, \tau)} = E_{i}\qty(\vb{x}, \tau). \label{App:EqnMotEffAction}
\end{equation}
In its Fourier form, this is the same equation as \req{eq:EqMotImTime} for a FE sample.

The linear susceptibility is related to the second derivative. We can easily verify the following identities for the second derivative
\begin{eqnarray}
    & \varepsilon_{0} \, \chi^{M}_{i, j}\qty(\vb{x}_{1}, \tau_{1}; \vb{x}_{2}, \tau_{2}) = \frac{\delta P_{i}\qty(\vb{x}_{1}, \tau_{1})}{\delta E_{j}\qty(\vb{x}_{2}, \tau_{2})} = -\frac{\delta^{2} W\qty[\vb{E}]}{\delta E_{i}\qty(\vb{x}_{1}, \tau_{1}) \, \delta E_{j}\qty(\vb{x}_{2}, \tau_{2})} \nonumber \\
    & = \qty[\frac{\delta E_{i}\qty(\vb{x}_{1}, \tau_{1})}{\delta P_{j}\qty(\vb{x}_{2}, \tau_{2})}]^{-1} = \qty[\frac{\delta^{2} \Gamma\qty[\vb{P}]}{\delta P_{i}\qty(\vb{x}_{1}, \tau_{1}) \, \delta P_{j}\qty(\vb{x}_{2}, \tau_{2})}]^{-1}. \label{App:SecFuncDerivs}
\end{eqnarray}
Using the chain rule, the third derivative \req{App:NonLinSuscDef} can be expressed as
\begin{eqnarray}
    & -\frac{\delta^{3} W\qty[\vb{E}]}{\delta E_{i}\qty(\vb{x}_{1}, \tau_{1}) \, \delta E_{j}\qty(\vb{x}_{2}, \tau_{2}) \, \delta E_{k}\qty(\vb{x}_{3}, \tau_{3})} \nonumber \\
    & = -\int d^{4}x_{6} \frac{\delta}{\delta P_{p}\qty(\vb{x}_{6}, \tau_{6})}\qty[\frac{\delta^{2} W\qty[\vb{E}]}{\delta E_{i}\qty(\vb{x}_{1}, \tau_{1}) \, \delta E_{j}\qty(\vb{x}_{2}, \tau_{2})}] \, \frac{\delta P_{p}\qty(\vb{x}_{6}, \tau_{6})}{\delta E_{k}\qty(\vb{x}_{3}, \tau_{3})} \nonumber \\
    & \stackrel{\mathrm{Eq.} \, \qty(\ref{App:SecFuncDerivs})}{=} \int d^{4}x_{6} \frac{\delta}{\delta P_{p}\qty(\vb{x}_{6}, \tau_{6})}\Bqty{\qty[\frac{\delta^{2} \Gamma\qty[\vb{P}]}{\delta P_{i}\qty(\vb{x}_{1}, \tau_{1}) \, \delta P_{j}\qty(\vb{x}_{2}, \tau_{2})}]^{-1}} \, \frac{\delta P_{p}\qty(\vb{x}_{6}, \tau_{6})}{\delta E_{k}\qty(\vb{x}_{3}, \tau_{3})} \nonumber \\
    & = -\int\int\int \dd{{}^{4}x_{4}} \dd{{}^{4}x_{5}} \dd{{}^{4}x_{6}} \qty[\frac{\delta^{2} \Gamma\qty[\vb{P}]}{\delta P_{i}\qty(\vb{x}_{1}, \tau_{1}) \, \delta P_{m}\qty(\vb{x}_{4}, \tau_{4})}]^{-1} \, \frac{\delta^{3} \Gamma\qty[\vb{P}]}{\delta P_{m}\qty(\vb{x}_{4}, \tau_{4}) \, \delta P_{n}\qty(\vb{x}_{5}, \tau_{5}) \, \delta P_{p}\qty(\vb{x}_{6}, \tau_{6})} \nonumber \\
    & \times \qty[\frac{\delta^{2} \Gamma\qty[\vb{P}]}{\delta P_{n}\qty(\vb{x}_{5}, \tau_{5}) \, \delta P_{j}\qty(\vb{x}_{2}, \tau_{2})}]^{-1} \, \frac{\delta P_{p}\qty(\vb{x}_{6}, \tau_{6})}{\delta E_{k}\qty(\vb{x}_{3}, \tau_{3})} \nonumber \\
    & \stackrel{\mathrm{Eq.} \, \qty(\ref{App:SecFuncDerivs})}{=} -\varepsilon^{3}_{0} \,  \int\int\int \dd{{}^{4}x_{4}} \dd{{}^{4}x_{5}} \dd{{}^{4}x_{6}} \chi^{M}_{i, m}\qty(\vb{x}_{1}, \tau_{1}; \vb{x}_{4}, \tau_{4}) \, \frac{\delta^{3} \Gamma\qty[\vb{P}]}{\delta P_{m}\qty(\vb{x}_{4}, \tau_{4}) \, \delta P_{n}\qty(\vb{x}_{5}, \tau_{5}) \, \delta P_{p}\qty(\vb{x}_{6}, \tau_{6})} \nonumber \\
    & \times \chi^{M}_{n, j}\qty(\vb{x}_{5}, \tau_{5}; \vb{x}_{2}, \tau_{2}) \, \chi^{M}_{p, k}\qty(\vb{x}_{6}, \tau_{6}; \vb{x}_{3}, \tau_{3}), \label{App:ThirdFuncDeriv},
\end{eqnarray}
where in the third line we used the matrix derivative identity
\[
\frac{\partial \hat{M}^{-1}}{\partial \lambda} = -\hat{M}^{-1} \cdot \frac{\partial \hat{M}}{\partial \lambda} \cdot \hat{M}^{-1}.
\]
Having in mind that the anharmonic free energy is local, the third variational derivative actually contains two Dirac delta functions
\begin{equation}
    \frac{\delta^{3} \Gamma\qty[\vb{P}]}{\delta P_{m}\qty(\vb{x}_{4}, \tau_{4}) \, \delta P_{n}\qty(\vb{x}_{5}, \tau_{5}) \, \delta P_{p}\qty(\vb{x}_{6}, \tau_{6})} = \frac{1}{\varepsilon_{0} \, P_{0}} \frac{\partial^{3}}{\partial \bar{P}_{m} \, \partial \bar{P}_{n} \, \partial \bar{P}_{p}}  \qty[\frac{\varepsilon_{0} \, \mathcal{F}_{\mathrm{ah}}\qty(\expval{\vb{P}\qty(\vb{x}_{4}, \tau_{4})})}{P^{2}_{0}}] \, \delta^{4}\qty(x_{5} - x_{4}) \, \delta^{4}\qty(x_{6} - x_{4}). \label{eq:ThirdDerivSimplify}
\end{equation}
Plugging \req{eq:ThirdDerivSimplify}, and expanding the susceptibilities in Fourier modes, assuming the developed polarization is uniform and does not break translation symmetry, we get
\begin{equation}
    \chi^{(2)}_{i, j, k}\qty(\vb{x}_{1}, \tau_{1}, \vb{x}_{2}, \tau_{2}, \vb{x}_{3}, \tau_{3}) = \qty(\frac{T}{L^{3}})^{2} \, \sum_{q_{1}, q_{2}} e^{-i \, \qty[ q_{1}\qty(x_{1} - x_{2}) + q_{2} \, \qty(x_{1} - x_{3})]} \chi^{(2)}_{i, j, k}\qty(q_{1}, q_{2}; \expval{\vb{P}}),
\end{equation}
\begin{equation}
    \chi^{(2)}_{i, j, k}\qty(q_{1}, q_{2}; \expval{\vb{P}}) = -\qty[\chi^{M}_{q_{1} + q_{2}, \mathrm{eff}}\qty(\expval{\vb{P}})]_{i, m} \, \frac{\partial^{3}}{\partial \bar{P}_{m} \, \partial \bar{P}_{n} \, \partial \bar{P}_{p}} \qty[\frac{\varepsilon_{0} \, \mathcal{F}_{\mathrm{ah}}\qty(\expval{\vb{P}})}{P^{2}_{0}}] \, \qty[\chi^{M}_{q_{1}, \mathrm{eff}}\qty(\expval{\vb{P}})]_{n, j} \, \qty[\chi^{M}_{q_{2}, \mathrm{eff}}\qty(\expval{\vb{P}})]_{p, k} \label{eq:NonLinSuscFourier}
\end{equation}
The Feynman diagram corresponding to \req{eq:NonLinSuscFourier} is presented in Fig.~\ref{fig:NonLinSusc}. For light, the wavevector is close to $\vb{q}_{1} = \vb{q}_{2} = \vb{0}$, and \req{eq:NonLinSuscFourier} is only a function of frequency. Because there is no frequency-dependent term in the third derivative (triangle vertes), it has a simple analytic continuation in the upper complex half-plane to get the retarded non-linear susceptibility
\begin{equation}
\chi^{\qty(2)}_{i, j, k}\qty(\omega_{1}, \omega_{2}; \expval{\vb{P}}) = - \qty[\chi^{\mathrm{ret}}_{\mathrm{eff}}\qty(\omega_{1} + \omega_{2}; \expval{\vb{P}})]_{i, m} \, \frac{\partial^{3}}{\partial \bar{P}_{m} \, \partial \bar{P}_{n} \partial P_{p}} 
    \qty[\frac{\varepsilon_{0} \, \mathcal{F}_{\mathrm{ah}}\qty(\expval{\vb{P}})}{P_{0}}] \, \qty[\chi^{\mathrm{ret}}_{\mathrm{eff}}\qty(\omega_{1}; \expval{\vb{P}})]_{n, j} \, \qty[\chi^{\mathrm{ret}}_{\mathrm{eff}}\qty(\omega_{2}, \expval{\vb{P}})]_{p, k}. \label{eq:NonLinSusc}
\end{equation}
\end{widetext}
\section{\label{App:DFT}DFT calculations}
We perform DFT calculations using the Projector Augmented-Wave method~\cite{paw} implemented in the Vienna Ab initio Simulation Package~\cite{vasp1, vasp2} version 6.3. The workflow management software atomate2~\cite{atomate2} is employed to organise and analyze the calculations. Our choice of pseudopotentials follows the standards of the materials project and atomate2. We select the Perdew-Burke-Ernzerhofer functional for solids~\cite{pbesol} as our exchange-correlation functional for its good balance between accurate geometries and relatively low cost~\cite{pbesol2}. 
A plane wave cutoff of $550$~eV is chosen together with automatically generated Gamma-centered $k$-point grids with a density of $50,000$ for the static calculations. We converge the static calculations to an energy difference of $1 \times 10^{-8}$~eV and forces during geometry optimizations to $1 \times 10^{-4}$~eV/\AA. All further calculation parameters can be found in the atomate2 json files provided in Ref.~\cite{materialscloud_entry}.

\section{\label{App:SSCHA} SSCHA calculations}

The SSCHA theory~\cite{sscha1,sscha2,sscha3,sscha4,sscha5} is designed to describe thermodynamics of crystal structures including temperature and quantum and anharmonic effects of the ions. The free energy at fininte temperature of the crystal structure is modelled in terms of the $N$-body density matrix of the ions. The latter is parameterized as a Gaussian in terms of average atomic positions and quantum and thermal fluctuations~\cite{sscha5}. The details of the implementation can be found in Ref.~\cite{sscha5}. 
Observables within SSCHA are calculated as Monte-Carlo averages, which require the calculation of ensembles of supercells of a crystal structure. We employ the force field described in Appendix~\ref{App:ML Force Field} to calculate these ensembles using an ensemble size of $4000$ $3 \times 3 \times 3$ supercells. The SSCHA minimization was run at $0 \, \mathrm{K}$, including quantum but not thermal effects, with default parameters besides a reduced minimization step size for the dynamical matrix (MIN\_STEP\_DYN=0.05). The structures were kept fixed to evaluate the energies at the desired polarization values.

\section{\label{App:ML Force Field}ML force fields}
 We employ Neural Equivariant Interatomic Potentials (NequIP) to model the STO Born-Oppenheimer energy surface. These are equivariant message passing networks and the details of their implementation can be found in Ref.~\cite{nequip}. The networks  were trained and tested using the data from Ref.~\cite{CesareSTO}. We split the training data from Ref.~\cite{CesareSTO} into a training set of 563 systems and a validation set of 63 systems. The network was trained until no further improvements in the loss could be found during the last 50 epochs.
 Using the same test set as discussed in the supplementary material of  Ref.~\cite{CesareSTO} we obtain energy and force RMSEs of $0.21$ meV/atom and $2$ meV/\AA{} vs $0.18$~meV/atom and $37$ meV/\AA{} in Ref.~\cite{CesareSTO}. This is an $18$-fold improvement in the force errors while retaining a comparable energy error. We used a cutoff radius of $6$~\AA, $4$ message passing steps, and a maximum order of $l = 2$ for the equivariant message passing. The input config and output files of NequIP including the model are available at Ref.~\cite{materialscloud_entry}.
 %will be uploaded to \url{https://doi.org/10.24435/materialscloud:7k-vk}.
 
 As the force field was trained using data created with slightly different DFT input parameters~\cite{CesareSTO} we confirmed that the fit parameters obtained using solely the force field are sufficiently similar to the ones obtained with the DFT parameters used in this work. Using the force field for the tetragonal cells we obtain $\alpha' = -2.8 \times 10^{8} \, \unitAlpha$, $\beta'_1 = 2.0 \times 10^{10} \, \unitBeta$ and $\beta'_2 = 0.57 \times 10^{10} \, \unitBeta$. While the double well becomes deeper with the force field the quartic coefficients and their ratio  stay nearly unchanged.

\section{Uncertainty in the fit parameters}
While the standard errors for the fit parameters are all at least one order of magnitude smaller than the parameters themselves, associating an exact uncertainty with the fit parameters is challenging, as many choices quantitatively influence the parameters. Examples include the direction of polarization chosen for the curves, the number of directions included in the fit, and whether the curves are calculated at constant or relaxed volume. Other factors are whether constant Born charges are assumed, and the choice of DFT parameters.  We demonstrate the impact of some of these choices and the robustness of the results with respect to them. 

Allowing for different volumes in the non-polar/polar structure and interpolating between them changes the ratio  $\beta'_1$/$\beta'_2$ to $2.5$ ($3.2$ at fixed volume) for the DFT calculations on the tetragonal structure and to $3.5$ ($4.0$ at fixed volume) for the SSCHA calculations.

We fitted the data in both polarization directions with a single linear fit. An alternative approach is to perform a separate fit of the [100] direction to obtain parameters for the quartic and square terms, followed by fitting the cross term for the [110] direction. For e.g., the cubic DFT calculations, this change yields $\alpha' = -0.94 \times 10^{8} \, \unitAlpha$, $\beta'_1 = 1.6 \times 10^{10} \, \unitBeta$, and $\beta'_2 = 0.44 \times 10^{10} \, \unitBeta$. This difference demonstrates that more data in different polarization directions would be optimal to obtain more accurate fitting parameters, however the difference is also sufficiently small to not influence the conclusions.

In a final test, we use different Born charges calculated with the atomate2 dielectric workflow for the non-polar tetragonal structure. 
This changes the fitting parameters to $\alpha' = -1.35 \times 10^{8} \, \unitAlpha$, $\beta'_1 = 1.4 \times 10^{10} \, \unitBeta$, and $\beta'_2 = 0.40 \times 10^{10} \, \unitBeta$.
\end{document}